\documentclass[a4paper,10pt]{article}
\usepackage[utf8]{inputenc}
\usepackage[a4paper, total={7in, 9in}]{geometry}

\usepackage{comment}
\usepackage{mathtools}
\usepackage{siunitx}
\usepackage{graphicx}
\usepackage{subcaption}
\usepackage{bbm}
\usepackage{pgfplots}
\usepackage{pgfplotstable}
\pgfplotsset{compat=1.3}
\usepackage{tikz}
\usepackage{authblk}
\usepgfplotslibrary{fillbetween}
\usetikzlibrary{patterns,calc,fit,plotmarks,tikzmark}

\usepackage{algorithm, algorithmicx, algpseudocode}

\usepackage{amsmath,amsbsy}
\usepackage[fleqn]{nccmath}
\usepackage{booktabs}
\usepackage{amsfonts}
\usepackage{amssymb}
\usepackage{caption}
\usepackage{cases}
\usepackage{comment}
\usepackage{graphicx}
\usepackage[colorlinks=true,bookmarksopen,bookmarksnumbered,linkcolor=black,citecolor=black,urlcolor=black]{hyperref}
\usepackage{mathtools}
\usepackage{moreverb}
\usepackage{multirow}

\usepackage{placeins}
\usepackage{tabularx}

\usepackage{multirow}
\usepackage{lmodern}
\usepackage{xparse}

\newcommand{\nd}{\vec{N}} 

\newcommand{\Hx}{\ensuremath{\mathbf{H}^\times}} 
\newcommand{\Bx}{\ensuremath{\mathbf{B}^\times}} 
\newcommand{\Ho}{\ensuremath{\mathbf{H}^\circ}} 
\newcommand{\Bo}{\ensuremath{\mathbf{B}^\circ}}
\providecommand{\keywords}[1]{\textbf{\textit{keywords---}} #1}

\usepackage{glossaries}  

\newacronym{cad}{CAD}{computer-aided design}
\newacronym[longplural=degrees of freedom]{dof}{DoF}{degree of freedom}
\newacronym{fe}{FE}{finite element}
\newacronym{fem}{FEM}{finite element method}
\newacronym{fit}{FIT}{finite integration technique}
\newacronym{nurbs}{NURBS}{non-uniform rational B-splines}
\newacronym{pec}{PEC}{perfect electric conductor}
\newacronym{pmc}{PMC}{perfect magnetic conductor}
\newacronym{rf}{RF}{radio-frequency}
\newacronym{te}{TE}{transverse electric}
\newacronym{tm}{TM}{transverse magnetic}
\newacronym{tem}{TEM}{transverse electric and magnetic}
\newacronym[longplural=boundary conditions]{bc}{BC}{boundary condition}

\newacronym{uq}{UQ}{uncertainty quantification}
\newacronym{pml}{PML}{perfectly matched layers}
\newacronym{fdtd}{FDTD}{finite-difference time-domain}
\newacronym{dgtd}{DGTD}{discontinuous Galerkin time-domain}
\newacronym{mim}{MIM}{metal-insulator-metal}
\newacronym{rhs}{RHS}{right-hand side}
\newacronym{sqp}{SQP}{sequential quadratic programming}
\newacronym{lar}{LAR}{least angle regression}
\newacronym{td}{TD}{total degree}
\newacronym{tp}{TP}{tensor product}
\newacronym{hc}{HC}{hyperbolic cross}
\newacronym{lhs}{LHS}{latin hypercube sampling}
\newacronym{ls}{LS}{least squares}
\newacronym{qmc}{QMC}{quasi Monte Carlo}
\newacronym{mlmc}{MLMC}{multilevel Monte Carlo}
\newacronym{hf}{HF}{high-frequency}
\newacronym{megpc}{MEGPC}{multi-element generalized polynomial chaos}
\newacronym{gp}{GP}{Gaussian process}
\newacronym{ed}{ED}{experimental design}
\newacronym{rms}{RMS}{root mean square}
\newacronym{cv}{CV}{cross-validation}
\newacronym{es}{ES}{educated sampling}
\newacronym{iga}{IGA}{isogeometric analysis}
\newacronym{bem}{BEM}{boundary element method}
\newacronym{ibvp}{IBVP}{initial/boundary value problem}

\newacronym{em}{EM}{electromagnetic}
\newacronym{emf}{EMF}{electromagnetic field}
\newacronym{gpc}{gPC}{generalized polynomial chaos}
\newacronym{mc}{MC}{Monte Carlo}
\newacronym{pde}{PDE}{partial differential equation}
\newacronym{pdf}{PDF}{probability density function}
\newacronym[longplural={quantities of interest}]{qoi}{QoI}{quantity of interest}
\newacronym{rv}{RV}{random variable}
\newacronym{ae}{a.e.}{almost every}

\newcommand{\mean}{\mathbb{E}}

\DeclareMathOperator*{\argmin}{\arg\!\min}

\newcommand{\drawTD}[2]
{
        \foreach \x in {0,1,2,...,#1} {
            \foreach \y in {0,1,2,...,#1} {
				\ifnum \numexpr\x+\y < \numexpr#1+1 
               		\node at (\x,\y) [circle, fill=#2, minimum size=5mm] {};
             	\fi 
            }
        }
}

\def\addlegendimage{\csname pgfplots@addlegendimage\endcsname}

\newcommand{\drawCoordinateSystem}[4] 
{

	\draw[thick,->] (#1+#4,0) -- (#2,0) node[anchor=south] {$z$};
	\draw[thick,->] (#1,#4) -- (#1,#3) node[anchor=north west] {$y$};
	\draw[thick] (#1,0) circle (#4);
	\draw[thick] (#1-0.707*#4,-0.707*#4) -- (#1+0.707*#4,0.707*#4){};
	\draw[thick] (#1-0.707*#4,0.707*#4) -- (#1+0.707*#4,-0.707*#4)node[anchor=south west] {$x$};
}

\pgfmathdeclarefunction{gauss}{2}{%
  \pgfmathparse{1/(#2*sqrt(2*pi))*exp(-((x-#1)^2)/(2*#2^2))}%
}

\tikzset{%
	remember picture with id/.style={%
		remember picture,
		overlay,
		save picture id=#1,
	},
	save picture id/.code={%
		\edef\pgf@temp{#1}%
		\immediate\write\pgfutil@auxout{%
			\noexpand\savepointas{\pgf@temp}{\pgfpictureid}}%
	},
	if picture id/.code args={#1#2#3}{%
		\@ifundefined{save@pt@#1}{%
			\pgfkeysalso{#3}%
		}{
			\pgfkeysalso{#2}%
		}
	}
}

\def\savepointas#1#2{%
	\expandafter\gdef\csname save@pt@#1\endcsname{#2}%
}

\def\tmk@labeldef#1,#2\@nil{%
	\def\tmk@label{#1}%
	\def\tmk@def{#2}%
}

\tikzdeclarecoordinatesystem{pic}{%
	\pgfutil@in@,{#1}%
	\ifpgfutil@in@%
	\tmk@labeldef#1\@nil
	\else
	\tmk@labeldef#1,(0pt,0pt)\@nil
	\fi
	\@ifundefined{save@pt@\tmk@label}{%
		\tikz@scan@one@point\pgfutil@firstofone\tmk@def
	}{%
		\pgfsys@getposition{\csname save@pt@\tmk@label\endcsname}\save@orig@pic%
		\pgfsys@getposition{\pgfpictureid}\save@this@pic%
		\pgf@process{\pgfpointorigin\save@this@pic}%
		\pgf@xa=\pgf@x
		\pgf@ya=\pgf@y
		\pgf@process{\pgfpointorigin\save@orig@pic}%
		\advance\pgf@x by -\pgf@xa
		\advance\pgf@y by -\pgf@ya
	}%
}

\definecolor{TUDa-0d}{cmyk/RGB/HTML}{0,0,0,.8/83,83,83/535353}
\definecolor{TUDa-0c}{cmyk/RGB/HTML}{0,0,0,.6/137,137,137/898989}
\definecolor{TUDa-0b}{cmyk/RGB/HTML}{0,0,0,.4/181,181,181/B5B5B5}
\definecolor{TUDa-0a}{cmyk/RGB/HTML}{0,0,0,.2/220,220,220/DCDCDC}

\definecolor{TUDa-1a}{cmyk/RGB/HTML}{.7,.4,0,0/93,133,195/5D85C3}
\definecolor{TUDa-2a}{cmyk/RGB/HTML}{0.8,.2,0,0/0,156,218/009CDA}
\definecolor{TUDa-3a}{cmyk/RGB/HTML}{0.7,0,.5,0/80,182,149/50B695}
\definecolor{TUDa-4a}{cmyk/RGB/HTML}{.4,0,.8,0/175,204,80/AFCC50}
\definecolor{TUDa-5a}{cmyk/RGB/HTML}{.2,0,.8,0/221,223,72/DDDF48}
\definecolor{TUDa-6a}{cmyk/RGB/HTML}{0,.1,.7,0/255,224,92/FFE05C}
\definecolor{TUDa-7a}{cmyk/RGB/HTML}{0,.3,.8,0/248,186,60/F8BA3C}
\definecolor{TUDa-8a}{cmyk/RGB/HTML}{0,.6,.8,0 /238,122,52/EE7A34}
\definecolor{TUDa-9a}{cmyk/RGB/HTML}{0,.8,.7,0/233,80,62/E9503E}
\definecolor{TUDa-10a}{cmyk/RGB/HTML}{.2,.9,0,0/201,48,142/C9308E}
\definecolor{TUDa-11a}{cmyk/RGB/HTML}{.6,.8,0,0/128,69,151/804597}
\definecolor{TUDa-1b}{cmyk/RGB/HTML}{1,.6,0,0/0,90,169/005AA9}
\definecolor{TUDa-2b}{cmyk/RGB/HTML}{1,.3,0,0/0,131,204/0083CC}
\definecolor{TUDa-3b}{cmyk/RGB/HTML}{1,0,.6,0/0,157,129/009D81}
\definecolor{TUDa-4b}{cmyk/RGB/HTML}{.5,0,1,0/153,192,0/99C000}
\definecolor{TUDa-5b}{cmyk/RGB/HTML}{.3,0,1,0/201,212,0/C9D400}
\definecolor{TUDa-6b}{cmyk/RGB/HTML}{0,.2,1,0/253,202,0/FDCA00}
\definecolor{TUDa-7b}{cmyk/RGB/HTML}{0,.4,1,0/245,163,0/F5A300}
\definecolor{TUDa-8b}{cmyk/RGB/HTML}{0,.7,1,0/236,101,0/EC6500}
\definecolor{TUDa-9b}{cmyk/RGB/HTML}{0,1,.9,0/230,0,26/E6001A}
\definecolor{TUDa-10b}{cmyk/RGB/HTML}{.4,1,0,0/166,0,132/A60084}
\definecolor{TUDa-11b}{cmyk/RGB/HTML}{.7,1,0,0/114,16,133/721085}
\definecolor{TUDa-1c}{cmyk/RGB/HTML}{1,.7,.2,0/0,78,138/004E8A}
\definecolor{TUDa-2c}{cmyk/RGB/HTML}{1,.5,.2,0/0,104,157/00689D}
\definecolor{TUDa-3c}{cmyk/RGB/HTML}{1,.2,.6,0/0,136,119/008877}
\definecolor{TUDa-4c}{cmyk/RGB/HTML}{.6,.1,1,0/127,171,22/7FAB16}
\definecolor{TUDa-5c}{cmyk/RGB/HTML}{.4,.1,1,0/177,189,0/B1BD00}
\definecolor{TUDa-6c}{cmyk/RGB/HTML}{.2,.3,1,0/215,172,0/D7AC00}
\definecolor{TUDa-7c}{cmyk/RGB/HTML}{.2,.5,1,0/210,135,0/D28700}
\definecolor{TUDa-8c}{cmyk/RGB/HTML}{.2,.8,1,0/204,76,3/CC4C03}
\definecolor{TUDa-9c}{cmyk/RGB/HTML}{.3,1,.9,0/185,15,34/B90F22}
\definecolor{TUDa-10c}{cmyk/RGB/HTML}{.5,1,.3,0/149,17,105/951169}
\definecolor{TUDa-11c}{cmyk/RGB/HTML}{.8,1,.2,0/97,28,115/611C73}
\definecolor{TUDa-1d}{cmyk/RGB/HTML}{1,.9,.3,0/36,53,114/243572}
\definecolor{TUDa-2d}{cmyk/RGB/HTML}{1,.7,.4,0/0,78,115/004E73}
\definecolor{TUDa-3d}{cmyk/RGB/HTML}{1,.4,.7,0/0,113,94/00715E}
\definecolor{TUDa-4d}{cmyk/RGB/HTML}{.7,.3,1,0/106,139,55/6A8B22}
\definecolor{TUDa-5d}{cmyk/RGB/HTML}{.5,.2,1,0/153,166,4/99A604}
\definecolor{TUDa-6d}{cmyk/RGB/HTML}{.4,.4,1,0/174,142,0/AE8E00}
\definecolor{TUDa-7d}{cmyk/RGB/HTML}{.3,.6,1,0/190,111,0/BE6F00}
\definecolor{TUDa-8d}{cmyk/RGB/HTML}{.4,.8,1,0/169,73,19/A94913}
\definecolor{TUDa-9d}{cmyk/RGB/HTML}{.5,1,.9,0/156,28,38/961C26}
\definecolor{TUDa-10d}{cmyk/RGB/HTML}{.7,1,.5,0/115,32,84/732054}
\definecolor{TUDa-11d}{cmyk/RGB/HTML}{.9,1,.3,0/76,34,106/4C226A}

\newcommand{\agcomment}[1]{\textcolor{black}{#1}}

\title{Data-Driven Solvers for Strongly Nonlinear Material Response}

\author{Armin Galetzka$^1$, Dimitrios Loukrezis$^{1,2}$, Herbert De Gersem$^{1,2,3}$}

\date{
	\small{$^1$Technische Universit\"at Darmstadt, Institute for Accelerator Science and Electromagnetic Fields (TEMF) \\ Schlossgartenstrasse 8, 64289 Darmstadt, Germany} \\ 
	\small{$^2$Technische Universit\"at Darmstadt, Centre for Computational Engineering \\ Dolivostrasse 15, 64293 Darmstadt, Germany} \\ 
	\small{$^3$Katholieke Universiteit Leuven - Kulak, Wave Propagation and Signal Processing Research Group \\ Etienne Sabbelaan 53, 8500 Kortrijk, Belgium}
}

\begin{document}
\maketitle
\begin{abstract}
	This work presents a data-driven magnetostatic finite-element solver that is specifically well-suited to cope with strongly nonlinear material responses.
	The data-driven computing framework is essentially a multiobjective optimization procedure matching the material operation points as closely as possible to given material data while obeying Maxwell's equations. Here, the framework is extended with heterogeneous (local) weighting factors - one per finite element - equilibrating the goal function locally according to the material behavior. This modification allows the data-driven solver \agcomment{to cope with unbalanced measurement data sets, i.e. data sets suffering from unbalanced space filling. This occurs particularly in the case of strongly nonlinear materials, which constitute problematic cases that hinder the efficiency and accuracy of standard data-driven solvers with a homogeneous (global) weighting factor.}
	The local weighting factors are embedded in the distance-minimizing data-driven algorithm used for noiseless data, likewise for the maximum entropy data-driven algorithm used for noisy data. 
	Numerical experiments based on a quadrupole magnet model with a soft magnetic material show that the proposed modification results in major improvements in terms of solution accuracy and solver efficiency.
	For the case of noiseless data, local weighting factors improve the convergence of the data-driven solver by orders of magnitude. 
	When noisy data are considered, the convergence rate of the data-driven solver is doubled.
	~\\~\\
\noindent\keywords{data-driven computing, data science, electromagnetic field simulation, noisy measurements, nonlinear material response, soft magnetic materials}
\end{abstract}
\maketitle

\section{Introduction}
Numerical simulations play an important role in the study of \gls{em} phenomena and are irreplaceable when designing \gls{em} devices.
Irrespective of the numerical approximation method of choice, e.g. the \gls{fem} \cite{jin2014, monk2003finite}, the \gls{bem} \cite{buffa2003galerkin}, or the \gls{fit} \cite{weiland1996time, weiland2003finite}, to name but a few, the underlying \glspl{ibvp} are governed by Maxwell's equations, which relate \gls{em} fields to their sources, and by constitutive laws, which describe material responses.
With respect to the latter, in the context of this work, we focus on the relation between the magnetic field strength $\vec{H}$ and the magnetic flux density $\vec{B}$, which depend on the magnetic permeability tensor $\boldsymbol{\mu}$, respectively, on the reluctivity tensor $\boldsymbol{\nu} = \boldsymbol{\mu}^{-1}$.
With the exception of the often irreducible aleatory uncertainties \cite{sullivan2015introduction}, Maxwell's equations are accepted as exact.
This is not true for the constitutive laws, which are in nearly all cases too complicated to be exactly described. 
For some simple materials such as vacuum, the constitutive relation is regarded as exactly known and given by $\vec{B}=\boldsymbol{\mu}\vec{H}$, $\boldsymbol{\mu}$ being a constant factor or tensor.
However, for most magnetic materials, the relation between $\vec{H}$ and $\vec{B}$ is strongly nonlinear and not exactly known.
In such cases, a constitutive law typically takes the form of an empirical model, which is most commonly based on fitting an analytical expression to experimental observation data.

The dominant paradigm in numerical \gls{em} computations is thus based on material models derived after available data.
These models are subsequently  provided to numerical \gls{em} field solvers \cite{brauer1975, heise1994}.
For that reason, significant effort is placed on improving the data-fitting and regression techniques that are used for material modeling purposes.
For example, advanced machine learning methods are used in an effort to capture complicated constitutive laws more accurately \cite{veeramani2009}. 
The modeling procedure can be further enhanced with constraints that the material model must conform to, based on physical considerations or domain expertise, \cite{gupta2015structure, pechstein2006}.
If large data sets are available, other approaches advocate to deduce material behavior purely based on the data \cite{rajan2013informatics, rajan2015materials}.
Nevertheless, regardless of implementation specifics and level of sophistication, data-based constitutive laws are uncertain in terms of the actual model describing them, a type of uncertainty called epistemic \cite{sullivan2015introduction}.
This model-form uncertainty unavoidably affects the solution of the numerical solver, thus rendering its predictions possibly questionable.
Moreover, the quality of these predictions cannot be improved with a finer spatial or temporal resolution, as the material model is a ``hard-coded'' element of the solver.

An alternative and fundamentally different paradigm regarding the incorporation of material data into numerical computations, bearing the name ``data-driven computing'', emerged recently from the field of computational mechanics.
This approach was introduced in the seminal work of Kirchdoerfer and Ortiz \cite{kirchdoerfer2016data} and was then further developed and analyzed for a range of problems arising in computational mechanics \cite{carrara2020data, conti2018data, conti2020data, eggersmann2019model, kirchdoerfer2018data, leygue2018data, stainier2019model}. For the case of \gls{em} simulations, a comparable formulation based on the variational principle was already proposed by Rikabi et al. \cite{rikabi1988} in 1988.
The main idea behind the data-driven computing paradigm is the reformulation of the \gls{ibvp} describing the physical phenomenon under investigation, such that the material modeling step is bypassed altogether, thus eliminating the attached epistemic uncertainty.
In this framework, the corresponding data-driven numerical solver operates directly on raw material data, essentially being material-model-free.
In particular, the data-driven solver seeks to minimize the distance between the field states in phase space fulfilling exact laws and the field states in phase space representing the constitutive law, where the latter is described in the form of a measurement data set.
In the present work, the term ``field state'' refers to a $\vec{B}$-$\vec{H}$ pair.

By relying on raw material data, the distance-minimization scheme is naturally sensitive to outliers in the data set.
Such outliers appear, for example, when we consider the case of noisy material data \cite{ayensa2018new}. 
The possibly dominant influence of the outliers has a significant impact on the robustness of data-driven solutions. 
To address this problem, the distance-minimization scheme can be modified, such that the data are first organized in clusters based on the maximum entropy principle and then the solver minimizes the free energy in phase space \cite{kirchdoerfer2017data}.
The data-driven computing paradigm has been further extended for the purpose of material response identification \cite{leygue2018data}, in which case the data-driven solver is effectively ``inverted'' as to locate states that sample the mechanical response of elastic materials.
A data-driven solver integrating the material response identification approach has also been developed, showing significant accuracy and efficiency improvements \cite{stainier2019model}. 
\agcomment{Approaches based on manifold learning are also available \cite{kanno2020kernel, ibanez2018manifold}. Thereby, the data-driven solution is not calculated using the original measurement data, but along a reduced-order manifold derived after the dataset. The manifold is constructed either locally, by means of local linear embedding \cite{ibanez2018manifold}, or globally, using kernel methods \cite{kanno2020kernel}.}
Outside the field of computational mechanics, data-driven magnetostatic \gls{fem} solvers were developed by the authors and co-workers, for the case of noise-free data and heterogeneous domains where exact and data-based material models co-exist \cite{degersem2020magnetic}.

The data-driven solvers \agcomment{which follow the original formulation \cite{kirchdoerfer2016data}}, employ a global weighting factor, thus treating all finite elements in the computational domain equally, regardless their material filling. 
However, a global, homogeneous factor may actually hinder the efficiency and accuracy of the data-driven solver in the case of \agcomment{incompatible material data sets, i.e. data sets where a solution in the material set is distant from the states that conform to the governing equations. Such material data sets are likely to occur in the case of strongly nonlinear materials, due to the fact that the nonlinear material response can result in dense sampling for some regions on the material curve, while others are only sparsely sampled. As a result, the data set does not accurately represent the underlying material law.}

The present work extends the data-driven computing paradigm in order to accommodate such strongly nonlinear material behavior as well.
Our contribution is motivated by the response of soft magnetic materials, the $BH$-curves of which typically consist of a very steep linear part, followed by the saturation part after a sharp transition.
To address the problem, we modify the data-driven solver's algorithm as to assign local weighting factors, i.e. one per finite element. 
Since no a priori knowledge about the working points along the $BH$-curve with respect to each finite element is available, the assignment of local weighting factors is performed adaptively during the data-driven solver's iterations.
We show that the proposed modification to the data-driven solver results in significant computational gains for noiseless and noisy material data alike. 
In the noiseless case, the data-driven solver utilizing local weighting factors converges orders of magnitude faster than the data-driven solver based on global factors.
For the case of noisy data, the convergence rate is improved by a factor greater than $2$.
Since data-driven methods based on both global and local weighting factors are applied and compared against each other, a side contribution of this work is the extension of previously introduced data-driven magnetostatic solvers \cite{degersem2020magnetic} to the case of noisy material data. 
Note that, with the introduction of local weighting factors, two previously suggested intrusive data-driven magnetostatic solvers \cite{degersem2020magnetic} are rendered unnecessary.

The rest of this paper is structured as follows. In Section~\ref{sec:general} the main idea behind the data-driven paradigm is described. Section~\ref{subsec:example} presents an illustrative example that showcases the data-driven computing framework. In Section~\ref{subsec:ms_dd} we specialize the data-driven formulation to the case of magnetostatics, which is then converted into the weak formulation in Section~\ref{subsec:weak_form} and subsequently discretized in Section~\ref{subsec:discrete_dd}. In Section~\ref{sec:sec_exactly_known_material} we recall the case of problems with heterogeneous domains, where both exactly known materials and data-based materials coincide. Lastly, Section~\ref{subsec:2D_weak_formulation} introduces the 2D data-driven weak formulation.
Section~\ref{sec:noisefree} is dedicated to the management of noiseless data. In Section~\ref{subsec:global_mat} we recall the usage of a global weighting factor over the entire domain, whereas in Section~\ref{subsec:local_mat} we present the novel approach with local weighting factors. In Section~\ref{subsec:noisefree_numerical_experiments} numerical experiments are carried out for both approaches on a 2D quadrupole magnet and show considerable improvements when local weighting factors are employed.
The handling of noisy measurement data is described in Section~\ref{sec:noisy}. First, we recall the basic idea of noisy measurement treatment introduced by Kirchdoerfer and Ortiz \cite{kirchdoerfer2017data} in Section~\ref{subsec:noisy_global}. Afterwards, we embed the local weighting factors into the framework of noisy measurement data. Numerical experiments show the enhancement brought by local weighting factors in Section~\ref{subsec:results_noisy}. \agcomment{A brief discussion on the computational complexity of the proposed algorithms is available in Section~\ref{sec:comp_complex}, followed by our concluding remarks in Section~\ref{sec:conclusion}.}

\section{Data-Driven Magnetostatic Solvers}
\label{sec:general}
\subsection{An introductory example}
\label{subsec:example}

\begin{figure}[t]
	\begin{subfigure}[t]{0.48\textwidth}
	\includegraphics[width=1.0\textwidth]{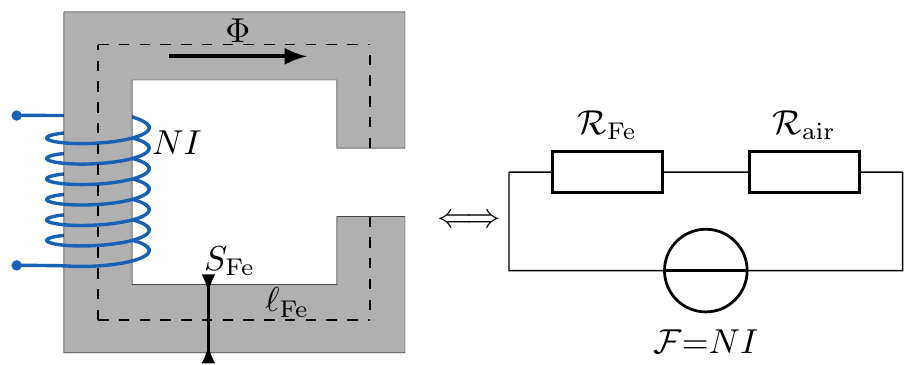} 
	\caption{}
	\label{fig:yoke}
	\end{subfigure}
	\hfill
	\begin{subfigure}[t]{0.48\textwidth}
	\centering
	\includegraphics[width=0.8\textwidth]{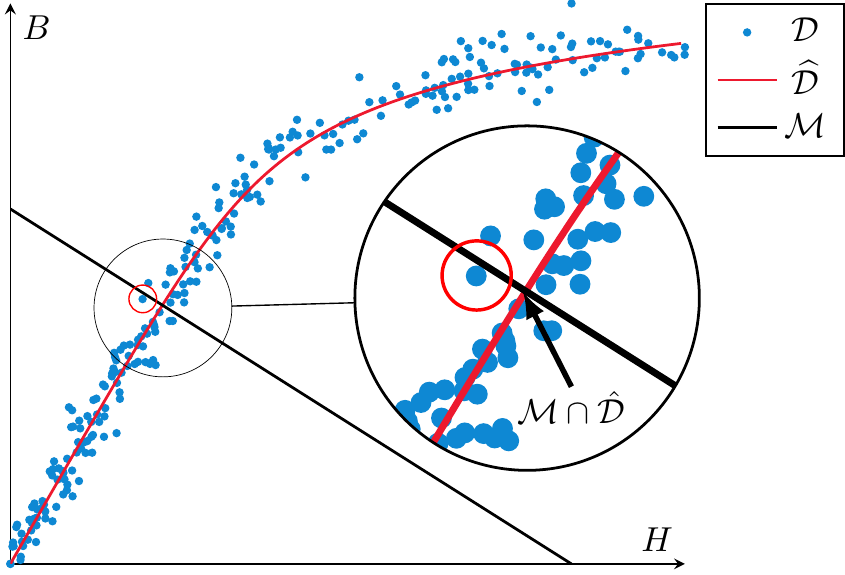} 
	\caption{}
	\label{fig:yoke_BH}
	\end{subfigure}		
	\caption{(a) C-yoke and equivalent magnetic circuit. (b) The black line shows the constrained set $\mathcal{M}$ of the states that fulfill the circuit law. The classical solution is given as the intersection of $\mathcal{M}$ and the set $\hat{\mathcal{D}}$, which corresponds to a regression-based $BH$-curve (shown in red). The data-driven solution is given by the state in the measurement data set closest to $\mathcal{M}$ (data given as blue dots, solution marked with a red circle).}
	\label{fig:yoke_and_BH}		
\end{figure}

Before deriving the general data-driven magnetostatic formulation, we first present an illustrative example of the data-driven computing paradigm within the context of magnetic field computations. 
In particular, we consider the C-shaped iron yoke depicted in Figure~\ref{fig:yoke}, where the magnetic flux $\Phi$ is generated by the current $I$ flowing through a coil with $N$ turns. \agcomment{Here and throughout the rest of this work, we consider the static case, i.e. all fields are in the steady state.}
Assuming no fringe fields on the outside of the iron core and a magnetic path with constant cross-section $S_\text{Fe}$, the field problem can be approximated with the equivalent magnetic circuit shown in Figure~\ref{fig:yoke}. Furthermore, on circuit level, it holds that $B = \left|\vec{B}\right|$ and $H = \left|\vec{H}\right|$, thereby assuming the magnetic field pairs $(H_\text{air},B)$ and $(H_\text{Fe},B)$, which are homogeneous in the air and iron regions, respectively.
The magnetomotive force $\mathcal{F}$ is then given as
\begin{equation}
\mathcal{F} = \oint \vec{H} \cdot \,\mathrm{d}\vec{s} = NI = \mathcal{F}_\text{Fe} + \mathcal{F}_\text{air}, 
\label{eq:magnetomotive_force}
\end{equation}
where $\mathcal{F}_\text{Fe}$, $\mathcal{F}_\text{air}$ refer to the magnetomotive forces in the iron yoke and in the air gap, respectively. Under the same assumptions, the magnetic field strength is constant, which allows us to express the magnetic flux as $\Phi = B S_\text{Fe}$. 
Hence, the magnetic flux can be expressed as
\begin{equation}
\Phi = \frac{\mathcal{F}}{\mathcal{R}_\text{Fe} + \mathcal{R}_\text{air}} = \frac{\mathcal{F}_\text{air}}{\mathcal{R}_\text{air}}= \frac{\mathcal{F}_\text{Fe}}{\mathcal{\mathcal{R}_\text{Fe}}},
\label{eq:magnetic_flux}
\end{equation}
where $\mathcal{R}_\text{Fe}$ and $\mathcal{R}_\text{air}$ refer to the magnetic reluctance of the yoke and the air gap, respectively. 
Combining \eqref{eq:magnetomotive_force} and \eqref{eq:magnetic_flux} results in
\begin{equation}
B S_\text{Fe} = \frac{-H_\text{Fe}\ell_\text{Fe}}{\mathcal{R}_\text{air}} + \frac{\mathcal{F}}{\mathcal{R}_\text{air}},
\label{eq:maxwell_example}
\end{equation}
where $\ell_\text{Fe}$ refers to the average length in the iron yoke, see Figure~\ref{fig:yoke}, and $H_\text{Fe}$ to the - currently unknown - magnetic field strength in the yoke. 

The goal is to find the state, i.e. the pair $(H_\text{Fe},B)$, that fulfills the circuit law \eqref{eq:maxwell_example}. 
All conforming states are collected in the set $\mathcal{M}$, such that
\begin{equation}
\mathcal{M} = \left\{ \zeta: \zeta = \left(H_\text{Fe},B\right) \in \mathcal{Z}: B S_\text{Fe} = \frac{-H_\text{Fe}\ell_\text{Fe}}{\mathcal{R}_\text{air}} + \frac{\mathcal{F}}{\mathcal{R}_\text{air}}\right\},
\label{eq:maxwell_states_example}
\end{equation}
where the set $\mathcal{Z}$ refers to all possible $(H,B)$ states.
\agcomment{For a given magnetomotive force $\mathcal{F}$}, all states that conform to the circuit law \eqref{eq:maxwell_example} form a line in the $(H_\text{Fe},B)$ phase space (black line in Figure~\ref{fig:yoke_BH}).
In order to find a solution to \eqref{eq:maxwell_example},  additional information regarding the relation between  $H_\text{Fe}$ and $B$ is necessary. 
This information is usually given in the form of measurement data, i.e. as discrete points $(H_{\text{Fe},m}, B_{m}), \, m=1,\dots,M,$ collected in a set $\mathcal{D}$ (blue points in Figure~\ref{fig:yoke_BH}). 
The traditional way of solving this problem involves fitting a continuous constitutive law to the given data, e.g. by means of regression (red line in Figure~\ref{fig:yoke_BH}).
The corresponding set of phase space states is then given by
\begin{equation}
	\hat{\mathcal{D}} = \left\{\zeta: \zeta = \left(H_\text{Fe},B\right) \in \mathcal{Z}: B=f\left(H_\text{Fe}\right)\right\},
\end{equation}
where $f(H_\text{Fe}): \mathbb{R}^+_0 \rightarrow \mathbb{R}^+_0$ defines the nonlinear $BH$-curve given by the red graph in Figure~\ref{fig:yoke_BH}.
The solution is then given as the intersection between the states fulfilling \eqref{eq:maxwell_states_example} and the states compatible with the constitutive law, i.e. $(H_\text{Fe},B) = \mathcal{M} \cap \hat{\mathcal{D}}$.
Diversely, in the data-driven paradigm, a solution in the intersection $\mathcal{M} \cap \mathcal{D}$ is sought. 

Note that the intersection between $\mathcal{M}$ and $\mathcal{D}$ is very likely to be an empty set, due to the fact that only a finite number of measurement data points are available.
Therefore, instead of searching for a state in $\mathcal{D}$, we accept as solution the state that fulfills the governing equation \eqref{eq:maxwell_example}, while at the same time being the closest to a state in $\mathcal{D}$. Accordingly, we search for the state that minimizes the distance between the governing equations and the available data set, given by
\begin{equation}
\zeta_\text{opt} = \argmin_{\zeta \in \mathcal{M}}{F\left(\zeta,\mathcal{D}\right)},
\label{eq:set_inter_relaxed}
\end{equation}
where $F(\zeta_1,\zeta_2)$ is a distance function between states $\zeta_1,\zeta_2$ in phase space.
This procedure is the core idea behind a data-driven solver, and is upscaled to a field formulation in the subsequent section.

\subsection{Magnetostatic data-driven formulation}
\label{subsec:ms_dd}
Next, we present the data-driven formulation for the magnetostatic case. 
The governing equations are Amp\`{e}re's law and Gauss's law for magnetism, which respectively read
\begin{subequations}
	\begin{alignat}{3}
	\mathrm{curl} \vec{H} &= \vec{J}, \quad &&\text{in~} \Omega, \label{eq:ampere}\\
	\mathrm{div} \vec{B} &= 0,        \quad &&\text{in~} \Omega, \label{eq:gauss}
	\end{alignat}
	\label{eq:magnetostatic}
\end{subequations}
where $\Omega$ denotes the computational domain \agcomment{and $\vec{J}$ the (source) current density.}
These equations are complemented with appropriate \glspl{bc} on the boundary of $\Omega$, denoted by $\Gamma$, thus leading to the ubiquitous setting of a boundary value problem.
We also assume that the two equations are free of any aleatory uncertainty, e.g. with respect to the geometry of the domain or the source terms. 
Aleatory uncertainty is of course possible and often inevitable, however, this case is not considered here. 

To solve the boundary value problem, a constitutive relation between $\vec{B}$ and $\vec{H}$ is necessary. 
In the traditional framework, a material model is derived after a measurement data set $\widetilde{\mathcal{D}}$ containing $(\vec{H},\vec{B})$-pairs for different operating points, such that
\begin{equation}
\widetilde{\mathcal{D}} = \left\{\left(\vec{H}^\star_1,\vec{B}^\star_1\right), \left(\vec{H}^\star_2,\vec{B}^\star_2\right), \dots, \left(\vec{H}^\star_M,\vec{B}^\star_M\right)\right\}.
\label{eq:measurement_set}
\end{equation}
In the context of data-driven solvers, the material modeling step is omitted. 
A formulation for both phase spaces is needed instead, i.e. regarding the phase space that covers the governing (exact) equations and the phase space addressing the material relation in view of the measurement set $\widetilde{\mathcal{D}}$.
Similar to Section~\ref{subsec:example}, the states that conform to the magnetostatic equations \eqref{eq:magnetostatic} can be found in the space
\begin{equation}
\agcomment{\mathcal{M} = \left\{ \zeta:  \zeta = \left(\vec{H},\vec{B}\right) \in \mathrm{H}(\text{curl}) \times \mathrm{H}(\text{div}), \\ 
\mathrm{curl} \vec{H}=\vec{J},\, \mathrm{div} \vec{B} = 0 \text{~and~\gls{bc}~} , \mathrm{a.e.} \in \Omega \right\}},
\label{eq:maxwell_set}
\end{equation}
where $\mathrm{H}(\text{curl})$ denotes the space of square-integrable functions with square-integrable curl and $\mathrm{H}(\text{div})$ the space of square-integrable functions with square-integrable divergence.
In order to derive the minimization problem as in \eqref{eq:set_inter_relaxed}, the local measurement set \eqref{eq:measurement_set} has to be reformulated such that the measurement set is valid in the entire domain of the underlying material, i.e. a spatial representation is necessary. Thus, we form a global set of discrete material states
\begin{equation}
\agcomment{\mathcal{D} = \left\{\zeta :\zeta \in  \mathrm{L}^2\times \mathrm{L}^2, \,\zeta(\vec{x}) \in \widetilde{\mathcal{D}},\,\mathrm{a.e.} \in \Omega\right\}.}
\label{eq:material_set}
\end{equation}
The data-driven solution is again given by the minimization problem \eqref{eq:set_inter_relaxed}. 
We thus accept as solution the state that conforms with the magnetostatic formulation and is best compatible with the available measurement data.

To properly calculate the distance between two states, a suitable norm in the phase space must be defined. 
That norm is given by 
\begin{equation}
\left|\left|\zeta\right|\right|^2_{\tilde{\boldsymbol{\mu}},\tilde{\boldsymbol{\nu}}} = \int_\Omega \left[\frac{1}{2} \widetilde{\boldsymbol{\mu}} \vec{H} \cdot \vec{H} + \frac{1}{2} \widetilde{\boldsymbol{\nu}} \vec{B} \cdot \vec{B} \right] \, \mathrm{d}\Omega,
\label{eq:distance}
\end{equation}
where $\zeta=(\vec{H},\vec{B})\in \mathcal{Z}=\mathrm{L}^2\times \mathrm{L}^2$ and  $\widetilde{\boldsymbol{\mu}}(\vec{x})$, $\widetilde{\boldsymbol{\nu}}(\vec{x}) \agcomment{= \widetilde{\boldsymbol{\mu}}(\vec{x})^{-1}}$ are tensorial and space-dependent weighting factors, respectively, and share units with the magnetic permeability and its inverse, the magnetic reluctivity, which explains the notation.
We emphasize that the weighting factors $\widetilde{\boldsymbol{\mu}}$ and  $\widetilde{\boldsymbol{\nu}}$ are of computational nature, i.e. they are not required to fulfill any physical material properties. 
Nevertheless, if appropriately chosen, they can improve the convergence of the data-driven solver significantly. 
\agcomment{In the magnetostatic case, the material tensors $\boldsymbol{\mu}$ and $\boldsymbol{\nu}$ can be reduced to diagonal matrices \cite{lacheisserie2005}. For anisotropic materials, the material properties are commonly measured along the main axes. It is always possible to locally apply rotation matrices such that the rotated coordinate system is aligned with the main axes of the material. Then, a diagonal tensor can be obtained. By that, the field components of $\vec{H}$ and $\vec{B}$ are decoupled and measurement data obtained from experiments \cite{arpaia2020, anglada2020}, can be directly incorporated in the data-driven solver. For that reason, we consider only diagonal weighting tensors for the rest of this work.} 
\agcomment{Furthermore, we consider weighting factors that satisfy} 
\begin{subequations}
\begin{alignat}{3}
&\agcomment{\mu_0} &&\le \agcomment{\widetilde{\boldsymbol{\mu}}(\vec{x})\cdot \vec{e}_d < \infty,} \\
&\agcomment{0} &&< \agcomment{\widetilde{\boldsymbol{\nu}}(\vec{x})\cdot \vec{e}_d \le \agcomment{\nu_0},}
\end{alignat}%
\label{eq:mu_constraints}%
\end{subequations}%
\agcomment{where $\mu_0, \nu_0$ denote the permeability and reluctivity in vacuum and $\vec{e}_d$ is the unit vector in the direction $d\in\{x,y,z\}$. 
This restriction is later demanded by the weak formulations to ensure well-posed problems.}

The distance function \agcomment{$F(\zeta,\zeta^\star)$} can now be written as
\begin{equation}
\agcomment{F(\zeta,\zeta^\star) 
= \int_\Omega f\left(\zeta,\zeta^\star\right)\,\mathrm{d}\Omega }
= \int_\Omega \left[\frac{1}{2} \left(\vec{H} - \vec{H}^\star\right)\cdot \widetilde{\boldsymbol{\mu}}  \left(\vec{H} - \vec{H}^\star\right) + \frac{1}{2} \left(\vec{B} - \vec{B}^\star\right) \cdot  \widetilde{\boldsymbol{\nu}}  \left(\vec{B} - \vec{B}^\star\right)\right]\,\mathrm{d}\Omega,
\label{eq:pen_func}
\end{equation}
where $\zeta^\star=(\vec{H}^\star,\vec{B}^\star)$ refers to the measurement data from the global set of discrete states defined in \eqref{eq:material_set}. \agcomment{Note that under the constraints \eqref{eq:mu_constraints}, the distance function \eqref{eq:pen_func} is convex.} With the chosen weighting factors, the distance function \eqref{eq:pen_func} returns the magnetic energy mismatch between two states.
The minimization problem \eqref{eq:set_inter_relaxed} can then be rewritten as
\begin{subequations}
\begin{alignat}{3}
&\text{minimize~}  &&\agcomment{F\left((\vec{H},\vec{B}),(\vec{H}^\star,\vec{B}^\star)\right)},			                 \label{eq:min_F} \\
&\text{subject~to~}&&\left\{\begin{array}{ll} \mathrm{curl}\vec{H} &= \vec{J},  \\
\mathrm{div}\vec{B} &= 0. \end{array}\right.
\label{eq:min_with_constraints}%
\end{alignat}%
\end{subequations}%
To fulfill Gauss's law \eqref{eq:gauss}, we introduce the magnetic vector potential $\vec{A}$, defined as $\vec{B} = \mathrm{curl} \vec{A}$. 
Now we can formulate a suitable ansatz to solve \eqref{eq:min_with_constraints}. 
The physical requirement for a divergence-free magnetic flux density $\vec{B}$ is fulfilled by incorporating the magnetic vector potential ansatz directly in \eqref{eq:min_F} and \eqref{eq:min_with_constraints}, whereas Amp\`{e}re's law \eqref{eq:ampere} is enforced by means of Lagrange multipliers. 
The stationary problem then reads
\begin{equation}
\mathcal{L}\left(\vec{H},\vec{A},\vec{\eta}\right)= F\left(\vec{H},\mathrm{curl} \vec{A}\right)   - G\left(\vec{\eta},\vec{H}\right),
\label{eq:lagrange_eq}
\end{equation}
where the term $G(\vec{\eta},\vec{H})$, which includes the Lagrange multiplier $\vec{\eta}(\vec{x})$, is given by
\begin{equation}
G\left(\vec{\eta},\vec{H}\right) = \int_\Omega \vec{\eta}\cdot \left(\vec{J} - \mathrm{curl} \vec{H}\right)\,\mathrm{d}\Omega.
\label{eq:lagrange_multi}
\end{equation}
Our goal is to find the stationary points of \eqref{eq:lagrange_eq}. 
To that end, we compute the functional derivative \cite{gelfand1963,weinstock1974} of \eqref{eq:lagrange_eq} with respect to $\vec{H},\vec{A}$, and $\vec{\eta}$. Since the functional derivative is linear, it is applied separately to \eqref{eq:pen_func} and \eqref{eq:lagrange_multi}.
Writing the distance function element-wise, we obtain
\begin{equation}\label{eq:distance_elementwise}
\begin{split}
F\left(\vec{H}, \mathrm{curl} \vec{A}\right) 
&= \int_\Omega  \frac{1}{2} \left[\widetilde{\mu}_x \left(H_x-H_x^\star\right)^2 + \widetilde{\mu}_y \left(H_y-H_y^\star\right)^2+\widetilde{\mu}_z \left(H_z-H_z^\star\right)^2 \right]\,\mathrm{d}\Omega   \\
&+ \int_\Omega \frac{1}{2} \left[\widetilde{\nu}_x \left(\partial_yA_z-\partial_zA_y-B_x^\star\right)^2  
+ \widetilde{\nu}_y \left(\partial_zA_x-\partial_xA_z-B_y^\star\right)^2 
+  \widetilde{\nu}_z \left(\partial_xA_y-\partial_yA_x-B_z^\star\right)^2 \right]\,\mathrm{d}\Omega,
\end{split}
\end{equation}
where $\partial_i \equiv \frac{\partial}{\partial_i}$ denotes the derivative with respect to the $i$-th component. 
The functional derivative of \eqref{eq:lagrange_multi} with respect to $\vec{A}$ is obviously vanishing. Then we must compute only the derivative of \eqref{eq:distance_elementwise}, which is given by
\begin{align}
\frac{\delta F}{\delta A_x} &= \frac{\partial f}{\partial A_x} - \partial_x \frac{\partial f}{\partial \{\partial_x A_x\}} - \partial_y \frac{\partial f}{\partial \{\partial_y A_x\}} - \partial_z \frac{\partial f}{\partial \{\partial_z A_x\}} \nonumber \\
&= \partial_y \left\{\widetilde{\nu}_z \left(\partial_xA_y-\partial_yA_x-B_z^\star\right) \right\}  - \partial_z \left\{\widetilde{\nu}_y \left(\partial_zA_x-\partial_xA_z-B_y^\star\right) \right\} 
\end{align}
\begin{align}
\frac{\delta F}{\delta A_y} &= \frac{\partial f}{\partial A_y} - \partial_x \frac{\partial f}{\partial \{\partial_x A_y\}} - \partial_y \frac{\partial f}{\partial \{\partial_y A_y\}} - \partial_z \frac{\partial f}{\partial \{\partial_z A_y\}} \nonumber \\
&=  \partial_z \left\{\widetilde{\nu}_x \left(\partial_yA_z-\partial_zA_y-B_x^\star\right) \right\}  - \partial_x \left\{\widetilde{\nu}_z \left(\partial_x A_y-\partial_yA_x-B_z^\star\right) \right\} 
\end{align}
\begin{align}
\frac{\delta F}{\delta A_z} &= \frac{\partial f}{\partial A_z} - \partial_x \frac{\partial f}{\partial \{\partial_x A_z\}} - \partial_y \frac{\partial f}{\partial \{\partial_y A_z\}} - \partial_z \frac{\partial f}{\partial \{\partial_z A_z\}} \nonumber \\
&=  \partial_x \left\{\widetilde{\nu}_y \left(\partial_zA_x-\partial_xA_z-B_y^\star\right) \right\}  - \partial_y \left\{\widetilde{\nu}_x \left(\partial_y A_z-\partial_zA_y-B_x^\star\right) \right\}, 
\end{align}
where the differential operator $\delta$ refers to the functional derivative.
Collecting all derivatives leads to
\begin{align}
\nabla_{\kern -0.2em A} F = \mathrm{curl}\left(\widetilde{\boldsymbol{\nu}}\mathrm{curl}\vec{A} - \widetilde{\boldsymbol{\nu}}\vec{B}^\star\right)
= \mathrm{curl}\left(\widetilde{\boldsymbol{\nu}}\mathrm{curl}\vec{A}\right) - \mathrm{curl}\left(\widetilde{\boldsymbol{\nu}}\vec{B}^\star\right),
\end{align} 
where $\nabla_{\kern -0.2em A}$ denotes the functional gradient with respect to the components of $\vec{A}$.
In the same manner, the functional derivative of \eqref{eq:lagrange_eq} with respect to the Lagrange multiplier $\vec{\eta}$ is given by 
\begin{equation}
\nabla_{\kern -0.2em \eta} \mathcal{L}=\vec{J} - \mathrm{curl}\vec{H}.
\end{equation}
Finally, the derivative of \eqref{eq:lagrange_eq} with respect to $\vec{H}$ is given as
\begin{align}
\nabla_{\kern -0.2em H} \mathcal{L}&= \nabla_{\kern -0.2em H}  F - \nabla_{\kern -0.2em H} G \nonumber \\
&= \widetilde{\boldsymbol{\mu}}(\vec{H} - \vec{H}^\star)+\mathrm{curl}\vec{\eta}.
\end{align}
Setting the derivatives to zero in order to solve for the stationary points of \eqref{eq:lagrange_eq}, we arrive at

\begin{subequations}
	\begin{alignat}{4}
	&\nabla_{\kern -0.2em A} &\mathcal{L} &: \quad &\mathrm{curl}\left(\widetilde{\boldsymbol{\nu}}\mathrm{curl}\vec{A}\right) - \mathrm{curl}\left(\widetilde{\boldsymbol{\nu}}\vec{B}^\star\right) &= 0, \label{eq:dA}\\ 
	&\nabla_{\kern -0.2em H}  &\mathcal{L}&:   \quad  &\widetilde{\boldsymbol{\mu}}(\vec{H} - \vec{H}^\star)+\mathrm{curl}\vec{\eta} &= 0, \label{eq:dH}\\
	&\nabla_{\kern -0.2em \eta} &\mathcal{L}&:\quad  &\vec{J} - \mathrm{curl}\vec{H} &= 0. \label{eq:deta}
	\end{alignat} 
	\label{eq:lagragne_derivative}
\end{subequations}

\subsection{Weak data-driven formulation}
\label{subsec:weak_form}
To solve the stationary problem \eqref{eq:lagragne_derivative}, we employ the \glsfirst{fem}. 
Let us for now assume that a particular state from $\mathcal{D}$ has been identified as the closest to conform with the magnetostatic equations. 
We denote that state with $(\vec{H}^\times,\vec{B}^\times)$.

As commonly done in the setting of the \gls{fem}, to obtain the weak formulations, we first multiply equation \eqref{eq:dA} with the test functions $\vec{w} \in V$, where $V$ is to be determined. Integration over the domain $\Omega$ yields
\begin{equation}
\int_\Omega \mathrm{curl}\left(\widetilde{\boldsymbol{\nu}}\mathrm{curl}\vec{A}\right)\cdot \vec{w}\,\mathrm{d}\Omega = \int_\Omega \mathrm{curl}\left(\widetilde{\boldsymbol{\nu}}\vec{B}^\times\right) \cdot \vec{w}\, \mathrm{d}\Omega.
\end{equation}
Applying Green's formula results in
\begin{equation}
\int_\Omega \widetilde{\boldsymbol{\nu}}\mathrm{curl}\vec{A} \cdot \mathrm{curl}\vec{w}\,\mathrm{d}\Omega = \int_\Omega \widetilde{\boldsymbol{\nu}} \vec{B}^\times \cdot \mathrm{curl}\vec{w}\, \mathrm{d}\Omega + \int_{\Gamma} (\widetilde{\boldsymbol{\nu}}\mathrm{curl}\vec{A} \times \vec{n}) \cdot \vec{w}\,\mathrm{d}S - \int_{\Gamma} (\widetilde{\boldsymbol{\nu}}\vec{B}^\times \times \vec{n}) \cdot \vec{w}\,\mathrm{d}S,
\label{eq:weak_A}
\end{equation}
where $\vec{n}$ is the unit normal vector at $\Gamma$.
The boundary is then split into a Neumann part $\Gamma_\text{N}$ and a Dirichlet part $\Gamma_\text{D}$. For the rest of this paper, we consider only homogeneous boundary conditions, which means the Neumann conditions are naturally fulfilled, whereas the Dirichlet conditions are enforced strongly in the chosen function space. Hence, the boundary integral term vanishes.
Next, we apply the curl operator on \eqref{eq:dH} and insert equation \eqref{eq:deta}. 
Multiplying again with the test functions $\vec{w}$ and integrating over the domain $\Omega$ leads to
\begin{equation}
\int_\Omega \mathrm{curl} \left(\widetilde{\boldsymbol{\nu}} \mathrm{curl}\vec{\eta}\right)\cdot \vec{w}\,\mathrm{d}\Omega = \int_\Omega \left[\mathrm{curl}\vec{H}^\times\cdot \vec{w} - \vec{J} \cdot \vec{w}\right]\,\mathrm{d}\Omega,
\end{equation}
which, after applying Green's formula, results in
\begin{equation}
\int_\Omega \widetilde{\boldsymbol{\nu}} \mathrm{curl}\vec{\eta} \cdot \mathrm{curl}\vec{w}\,\mathrm{d}\Omega = \int_\Omega \left[\vec{H}^\times \cdot \mathrm{curl}\vec{w} - \vec{J} \cdot \vec{w}\right]\,\mathrm{d}\Omega + \int_{\Gamma}\left(\mathrm{curl}\vec{\eta}\times \vec{n}\right) \cdot \vec{w}\,\mathrm{d}S - \int_{\Gamma}\left(\vec{H}^\times \times \vec{n}\right) \cdot \vec{w}\,\mathrm{d}S.
\label{eq:weak_eta}
\end{equation}
We then apply the same boundary conditions as in \eqref{eq:weak_A}. 
For notation purposes, we introduce the inner product
\begin{equation}
(\vec{u},\,\vec{v})_\Omega = \int_\Omega \vec{u} \cdot \vec{v}\, \mathrm{d}\Omega,
\label{eq:inner_product}
\end{equation}
where $\vec{u},\, \vec{v} \in \mathrm{L}^2(\Omega)^3$.
Now, we can express \eqref{eq:weak_A}  and \eqref{eq:weak_eta} using the bilinear forms
\begin{subequations}
\begin{align}
a(\vec{A},\vec{w}) &:= (\widetilde{\boldsymbol{\nu}}\mathrm{curl}\vec{A},\mathrm{curl}\vec{w})_\Omega, \\
a(\vec{\eta},\vec{w}) &:= (\widetilde{\boldsymbol{\nu}}\mathrm{curl}\vec{\eta},\mathrm{curl}\vec{w})_\Omega,
\end{align}
\label{eq:bilinear_form}%
\end{subequations}
where the \glspl{rhs} are given by
\begin{subequations}
	\begin{align}
	l_1^\times(\vec{w})&=(\widetilde{\boldsymbol{\nu}} \vec{B}^\times,\mathrm{curl}\vec{w})_\Omega = \int_\Omega \widetilde{\boldsymbol{\nu}} \vec{B}^\times \cdot \mathrm{curl}\vec{w}\, \mathrm{d}\Omega \label{eq:rhs1}, \\
	l_2^\times(\vec{w})&= (\vec{H}^\times,\mathrm{curl}\vec{w})_\Omega - (\vec{J},\vec{w})_\Omega= \int_\Omega \left[\vec{H}^\times \cdot \mathrm{curl}\vec{w} - \vec{J} \cdot \vec{w}\right]\,\mathrm{d}\Omega. \label{eq:rhs2}
	\end{align}
	\label{eq:rhss}%
\end{subequations}
Note that the superscript $^\times$ on the functionals in \eqref{eq:rhs1} and \eqref{eq:rhs2} signifies that these functionals incorporate states given by the material data set. Also note that the \glspl{rhs} of \eqref{eq:rhss} are of non-standard type as they feature a singular excitation term through the measurement data.

In contrast to the traditional approach, the data-driven formulation requires the solution of two linear systems. 
However, the bilinear forms corresponding to \eqref{eq:weak_A} and \eqref{eq:weak_eta} are identical, therefore, the system matrix must be assembled only once.
The weak formulations then read: Find $\vec{A},\vec{\eta}\in V=\{\vec{v} \in \mathrm{H}(\text{curl}): \vec{v}\times \vec{n}=0 \text{~on~}\Gamma_\text{D}\}$ such that
\begin{subequations}
	\begin{align} 
	a(\vec{A},\vec{w})    &= l_1^\times(\vec{w}),  \quad \forall \vec{w}\in V, \\
	a(\vec{\eta},\vec{w}) &= l_2^\times(\vec{w}),  \quad \forall \vec{w}\in V.
	\end{align}
	\label{eq:weak_form_with_zero_kernel}%
\end{subequations}
Note that, in their current format, the problems \eqref{eq:weak_form_with_zero_kernel} are not well-posed. Both vector fields $\vec{A}$ and $\vec{\eta}$ are only defined up to a gradient, i.e. they feature a large kernel given by the gradients of  $\mathrm{H}(\mathrm{grad})$ functions. It follows, that the bilinear forms are not coercive, i.e. $a(\vec{u},\vec{u})\ngeq ||\vec{u}||^2_{\mathrm{H}(\mathrm{curl})}$. 
For instance, if $\vec{u}=\mathrm{grad}\Phi$, then it directly follows that $a(\vec{u},\vec{u})=0$, but also $||\vec{u}||^2_{\mathrm{H}(\mathrm{curl})}=||\mathrm{grad}\Phi||^2_{L^2}$. 
There are several strategies to ensure that \eqref{eq:weak_form_with_zero_kernel} is satisfied for all test functions, e.g. current vector potential formulations \cite{biro1993RHScurrentPotential, ren1996curlcurlRHS} to ensure a compatible \gls{rhs}, imposing the Coulomb gauge \agcomment{or, on a discrete level, co-tree gauging \cite{dular1995cotree, creuse2019cotree}}, to name but a few.

After solving \eqref{eq:weak_form_with_zero_kernel}, we obtain an updated field solution. The magnetic flux density is directly computed after the magnetic vector potential $\vec{A}$, whereas the magnetic field strength $\vec{H}$ is obtained by employing \eqref{eq:dH}. 
The updated terms read
\begin{subequations}
	\begin{align}
		\vec{B} &= \mathrm{curl}\vec{A}, \\
		\vec{H} &= \vec{H}^\times + \widetilde{\boldsymbol{\nu}} \mathrm{curl}\vec{\eta}.		
	\end{align}
	\label{eq:update_field}%
\end{subequations}
Next, we have to find the optimal states in $\mathcal{D}$, i.e. states $(\vec{H}^\star,\vec{B}^\star) \in \mathcal{D}$ that are closest to \eqref{eq:update_field}. 
In this manner, we find the states in the space of $(\vec{H},\vec{B})$-pairs that fulfill the constitutive relation, while at the same time are closest to the magnetostatic formulation. 
The optimal pairs are determined with 
\begin{equation}
\left(\vec{H}^\times,\vec{B}^\times\right) = \argmin_{(\vec{H}^\star,\vec{B}^\star) \in \mathcal{D}} \left\{ \agcomment{F\left((\vec{H},\vec{B}),(\vec{H}^\star,\vec{B}^\star)\right)}  \right\}.
\label{eq:semidescrite_min}
\end{equation}
These new optimal states are taken into account to solve \eqref{eq:weak_form_with_zero_kernel}, followed by updating the field solution \eqref{eq:update_field} and minimizing \eqref{eq:semidescrite_min}. 
These computations are performed iteratively, until there is no change in $(\vec{H}^\times,\vec{B}^\times)$ in two consecutive iterations. At the beginning, the ``optimal'' state is randomly initialized within the material data set $\mathcal{D}$.

The data-driven algorithm in continuous (weak) formulation is given in Algorithm~\ref{alg:dd_weak_form}.

	\begin{algorithm}[t!]
	\begin{algorithmic}
		\State \textbf{initialize} $(\vec{H}^\times, \vec{B}^\times)$ randomly on $\mathcal{D}$ 
		\While{convergence not reached}
			\vspace{0.5em}
			    \State Find~$ \vec{A},\vec{\eta} \in V$ such that
				\State \phantom{Update~~~~~~}	$a(\vec{A},\vec{w}) = l_1^\times(\vec{w}) \quad \forall \vec{w} \in V $   
				\State \phantom{Update~~~~~~}  $a(\vec{\eta},\vec{w})\, = l_2^\times(\vec{w})\quad \forall \vec{w} \in V$
				\State Update~~~~~~ $\vec{B} = \mathrm{curl}\vec{A}$
				\State \phantom{Update~~~~~~} $\vec{H}  = \vec{H}^\times + \widetilde{\boldsymbol{\nu}} \mathrm{curl}\vec{\eta}$
			\State Find $(\vec{H}^\times, \vec{B}^\times)$ in $\mathcal{D}$ adjacent to $(\vec{H}, \vec{B})$ with \eqref{eq:semidescrite_min}. 
		\EndWhile
	\end{algorithmic}
		\caption{Iterative scheme for the data-driven magnetostatic field solver in continuous (weak) formulation.}
		\label{alg:dd_weak_form}
	\end{algorithm}  

\subsection{Discrete data-driven formulation}
\label{subsec:discrete_dd}
For the rest of this work, lowest order \glspl{fe} are employed.
When discretizing the weak forms in \eqref{eq:weak_form_with_zero_kernel}, curl-conforming ansatz and test functions must be used for $\vec{A}$ and $\vec{\eta}$. The magnetic vector potential and the Lagrange multiplier are thus discretized as
\begin{subequations}
\begin{align}
\vec{A} \approx \vec{a}_h = \sum_{i=1}^{N_\text{e}} a_i \nd_i(\vec{x}), \\
\vec{\eta} \approx \vec{\eta}_h = \sum_{i=1}^{N_\text{e}} \eta_i \nd_i(\vec{x}),
\end{align}
\label{eq:discreteAandeta}%
\end{subequations}
where $\nd_i$ are N\'{e}d\'{e}lec basis functions of first kind defined on a triangulation of $\Omega$ \cite{nedelec1980}, the subscript $h$ denotes the maximum edge length of the triangulation, $N_\text{e}$ the number of \glspl{dof} allocated on edges, and $a_i$, $\eta_i$ the degrees of freedom. 
The Galerkin approach is used, such that ansatz and test functions are identical. The algebraic representation of the weak forms in \eqref{eq:weak_form_with_zero_kernel} read
\begin{subequations}
\begin{align}
\mathbf{K}_{\tilde{\boldsymbol{\nu}}} \mathbf{a} &= \mathbf{b}_1^\times \\
\mathbf{K}_{\tilde{\boldsymbol{\nu}}} \boldsymbol{\eta} &= \mathbf{b}_2^\times 
\end{align}
\label{eq:discrete_problem_definition}%
\end{subequations}
where
\begin{subequations}
	\begin{align}
	\mathbf{K}_{\tilde{\boldsymbol{\nu}},i,j} &= (\widetilde{\boldsymbol{\nu}} \mathrm{curl}\nd_j,\, \mathrm{curl} \nd_i)_\Omega, \label{eq:stiffnes_matrix}\\
	\mathbf{b}_{1,i}^\times &= (\widetilde{\boldsymbol{\nu}} \vec{B}^\times,\,\mathrm{curl}\nd_i)_\Omega, \label{eq:rhs_1_discrete}\\
	\mathbf{b}_{2,i}^\times &= (\vec{H}^\times,\,\mathrm{curl}\nd_i)_\Omega - (\vec{J},\nd_i)_\Omega. 
	\end{align} 
	\label{eq:discrete_operators_definition}%
\end{subequations}
and $\mathbf{a} \in \mathbb{R}^{N_\text{e}}$, respectively $\boldsymbol{\eta} \in \mathbb{R}^{N_\text{e}}$ are the \glspl{dof} of the corresponding vector fields in \eqref{eq:discreteAandeta}.
Again, $^\times$ indicates that material data was used to create the \gls{rhs}. Furthermore, we introduce the diagonal matrices $\mathbf{D}_\Omega$ as well as the matrices containing the weighting factors $\mathbf{D}_{\tilde{\boldsymbol{\nu}}}$, respectively $\mathbf{D}_{\tilde{\boldsymbol{\mu}}}$. The matrices are defined by
\begin{align}
\mathbf{D}_{\Omega,d,e,e} &= \int_{T_e}  \mathrm{d}\,\Omega, = V_e, \\
\mathbf{D}_{\tilde{\boldsymbol{\nu}},d,e,e} &=  \widetilde{\boldsymbol{\nu}}(\vec{x}_e), \label{eq:Dnu} \\ 
\mathbf{D}_{\tilde{\boldsymbol{\mu}},d,e,e} &=  \widetilde{\boldsymbol{\mu}}(\vec{x}_e),  \label{eq:Dmu}
\end{align}
where $d\in \{x,y,z\}$, $e \in \{1,\dots,N_\mathrm{elem}\}$, $T_e$ denotes the $e$-th element of the triangulation of $\Omega$ and $\vec{x}_e$ refers to the center point of the corresponding elements.
The discrete update terms are given by
\begin{subequations}
	\begin{align}
		\vec{B}_h &= \mathrm{curl}\vec{a}_h \\
		\vec{H}_h &= \vec{H}_h^\times + \widetilde{\boldsymbol{\nu}}\mathrm{curl}\vec{\eta}_h. 
	\end{align}
	\label{eq:discrete_update}%
\end{subequations}
We will denote the values of the fields in \eqref{eq:discrete_update} at the center points $\vec{x}_e$ by $\Bo$, respectively $\Ho$. 
\agcomment{The fields are evaluated at the quadrature points of the \glspl{fe} \cite{degersem2020magnetic}. Note that, in this work, we consider lowest-order \glspl{fe}, accordingly, one point per element.}
To obtain the discrete counterpart of the distance function \eqref{eq:pen_func} we introduce the matrix induced scalar products
\begin{align}
\langle \mathbf{H},\mathbf{H} \rangle_{\mathbf{D}_{\tilde{\boldsymbol{\nu}}}} &= \mathbf{H}^\top \mathbf{D}_{\Omega} \mathbf{D}_{\tilde{\boldsymbol{\nu}}}  \mathbf{H} , \\
\langle \mathbf{B},\mathbf{B} \rangle_{\mathbf{D}_{\tilde{\boldsymbol{\mu}}}} &= \mathbf{B}^\top \mathbf{D}_{\Omega} \mathbf{D}_{\tilde{\boldsymbol{\mu}}}  \mathbf{B} ,
\end{align}
for $\mathbf{H},\mathbf{B}\in \mathbb{R}^{N_\mathrm{elem}}$. Now, the discrete minimization problem reads 
\begin{align}
\left(\mathbf{H}^\times,\mathbf{B}^\times\right)&=\argmin_{(\mathbf{H}^\star,\mathbf{B}^\star)\in \mathcal{D}} F(\Ho, \Bo) \nonumber \\ 
&=  \argmin_{(\mathbf{H}^\star,\mathbf{B}^\star)\in \mathcal{D}} 
\frac{1}{2}\langle \Ho-\mathbf{H}^\star,\Ho-\mathbf{H}^\star  \rangle_{\mathbf{D}_{\tilde{\boldsymbol{\nu}}}} +\frac{1}{2} \langle \Bo-\mathbf{B}^\star,\Bo-\mathbf{B}^\star  \rangle_{\mathbf{D}_{\tilde{\boldsymbol{\mu}}}}. 
\label{eq:distance_discrete}
\end{align}
The distance minimization is carried out for each dimension separately.

\subsection{Data-driven solver with exactly known material relation}
\label{sec:sec_exactly_known_material}
In practical, real-world applications, it is not uncommon that some regions of the computational domain $\Omega$ consist of materials with exactly known properties, such that there exists a closed-form relation between $\vec{H}$ and $\vec{B}$. 
For example, this situation is encountered in models where iron parts and air gaps coexist.
A straightforward strategy is to provide a measurement data set $\mathcal{D}$ containing all possible (infinite) points that obey the closed-form relation.
The data-driven computing framework can then be modified accordingly. 
In that case, the weighting factor in the domain with known constitutive law should be chosen to be equal to the now known material coefficients.

A previous work of the authors proposes three modifications of increasing intrusiveness to the data-driven computing framework, in order to deal with mixed models, i.e. with exact and data-based materials co-existing in the computational domain \cite{degersem2020magnetic}.
In this work, we shall stick to the least intrusive approach, where the exact material relation is treated by the data-driven solver itself. 
Under this approach, the distance function \eqref{eq:distance_discrete} is updated with the known material coefficients $\boldsymbol{\mu}_\text{ex}$ and $\boldsymbol{\nu}_\text{ex}$, and the solution of the minimization problem is given by 
\begin{subequations}
	\begin{align}
	\Bx &= \frac{\Bo + \mathbf{D}_{\boldsymbol{\mu}_\text{ex}} \Ho}{2}, \\
	\Hx &= \mathbf{D}_{\boldsymbol{\nu}_\text{ex}}\Bx,
	\end{align}
	\label{eq:exactly_known_BH}%
\end{subequations}
where $\mathbf{D}_{\boldsymbol{\nu}_\text{ex}},\, \mathbf{D}_{\boldsymbol{\mu}_\text{ex}}$ are matrices following the definition in \eqref{eq:Dnu}, respectively \eqref{eq:Dmu}.
The field solution is then updated with \eqref{eq:exactly_known_BH} at the same spot as the field solution for the parts with unknown material relation.
Note that, during the assembly of $\mathbf{K}_{\tilde{\boldsymbol{\nu}}}$ and in order to  update the terms for the magnetic field strength, a mixed weighting factor has to be considered. 
Thus, the weighting factor $\widetilde{\boldsymbol{\nu}}$ now contains exactly known material coefficients for the domains with known material law and weighting factors for the purely data-based domains.

Other known material laws can be treated in the same manner. Nevertheless, the relations \eqref{eq:exactly_known_BH} only hold for the case of linear materials. 
For the nonlinear case, the corresponding relations have to be derived by minimizing \eqref{eq:distance_discrete} in view of the nonlinear material law.

\subsection{2D magnetostatic weak formulation}
\label{subsec:2D_weak_formulation}
In many engineering applications, it is possible to reduce the more accurate but computationally challenging three-dimensional formulation to a less demanding two-dimensional one. 
For example, such a reduction is possible if the geometry of the considered device remains almost unchanged along one certain direction. 
Consequently, the field component in that particular direction can be regarded as negligible.
In the 2D case, the magnetic vector potential is reduced to one spatial component. Without loss of generality, we choose the $z$-component, such that $\vec{A} = (0,0,A_z)$, respectively $\vec{\eta} = (0,0,\eta_z)$. 
The edge functions can then be chosen to be
\begin{equation}
	\nd_\text{2D} = \Phi(x,y)\vec{e}_z,
\label{eq:2dbasis}
\end{equation}
where $\Phi$ are standard nodal functions defined on a triangulation of the 2D cross-section of $\Omega$ and $\vec{e}_z$ denotes the unit vector in $z$-direction. Employing the 2D basis function \eqref{eq:2dbasis} as test and trial functions in \eqref{eq:weak_form_with_zero_kernel} leads to two linear systems that must be solved \cite{degersem2020magnetic}. Note that the constructed 2D basis function \eqref{eq:2dbasis} have zero divergence, meaning an additional gauging can be omitted.
It should also be pointed out that the structure of Algorithm~\ref{alg:dd_weak_form} and its discrete counterpart remain unchanged.

\section{Data-Driven Magnetostatic Simulation with Noiseless Data}
\label{sec:noisefree}
Until now, the choice of the weighting factors $\widetilde{\boldsymbol{\nu}}$ within the context of data-driven solutions has not been addressed.
In contrast to conventional \gls{fem} solvers, the factor $\widetilde{\boldsymbol{\nu}}$ does not necessarily represent the physical behavior of the material, but is rather an element of computational nature. 
Nevertheless, choosing the weighting factors in an appropriate way can increase the convergence rate of the data-driven algorithm substantially. 
In the following, we discuss two different possibilities, i.e. global and local weighting factors.
Further, we assume that the measurement pairs $(\vec{H}^\star_{m},\vec{B}^\star_{m}),\,m=1,\dots,M$ are noise-free, such that the observed values are unaffected by  measurement errors or other uncertainty sources.
The case of noisy data is discussed in Section~\ref{sec:noisy}.

The most naive choice for the weighting factors would be $\widetilde{{\nu}}=1$, respectively $\widetilde{{\mu}}=1$. However, calculating a distance in the $(H,B)$ phase space already requires stretching and squeezing the axes, because the values for $H$ and $B$ differ by orders of magnitude. Moreover, the ratio between $H$ and $B$ depends strongly on the material, thus necessitating heterogeneous weighting factors. The ratio between $H$ and $B$ is also strongly affected by the nonlinearity, which is our major concern here, and is counteracted by adaptively updating $\widetilde{\boldsymbol{\nu}}$ and $\widetilde{\boldsymbol{\mu}}$.

In the case of a nonlinear material, the constitutive law reads $H(B)=\nu(B)B$. At a most recently obtained state $(H^\circ,B^\circ)$, we can distinguish between two possible linearizations, corresponding to the successive-substitution method and the Newton method, respectively. The corresponding operations points read
\begin{align}
H-H^\circ &= \nu_\mathrm{c}^\circ (B-B^\circ), \\
H-H^\circ &= \nu_\mathrm{d}^\circ (B-B^\circ), 
\end{align}
where
\begin{align}
\nu_\mathrm{c}^\circ &= \frac{H^\circ}{B^\circ}, \\
\nu_\mathrm{d}^\circ &= \frac{\mathrm{d}H^\circ}{\mathrm{d}B^\circ}, 
\end{align}
are the \emph{chord} and \emph{differential} reluctivity, respectively \cite{degersem2008material} (see Figure~\ref{fig:BHmu}). The linearization is carried out for each \gls{fe} quadrature point and forces a reassembly of the \gls{fe} matrices in each iteration step. For the problems under consideration, up to $100$ successive-substitution iterations or up to $20$ Newton iterations are needed to obtain convergence, depending on the desired accuracy.

The data-driven solver does not dispose of a material law and, thus, does not inherently incorporate an iteration for treating the nonlinearity. One could consider the data-driven iteration as a replacement thereof. The updates \eqref{eq:discrete_update} and \eqref{eq:distance_discrete} then correspond to the action of linearization in a traditional solver. Hence, it may not wonder that particular choices for $\widetilde{\boldsymbol{\nu}}$ and $\widetilde{\boldsymbol{\mu}}$ proposed below, show similarities to the standard successive-substitution and Newton methods. 

For the rest of this work, we determine the weighting factors according to the differential reluctivity. Note that one could also consider the chord reluctivity, which still results in an improved convergence compared to the original method. However, the results cannot compete with the ones obtained when employing the differential reluctivity. Also note that, when global weighting factors are employed, the results obtained using the chord or the differential reluctivity differ only marginally.

\subsection{Global weighting factor}
\label{subsec:global_mat}
A straightforward method to get an estimation of the global weighting factor $\widetilde{\boldsymbol{\nu}}$, is to apply finite differences on the measurement data $\mathcal{D}$ to obtain the differential reluctivity. 
We consider here the case of anisotropic materials and decoupled fields, such that $H_d$ depends only on $B_d$, where ${d} \in \{x,y,z\}$ refers to the spatial dimension. 
Then, the reluctivity tensor has only diagonal elements and reads
\begin{equation}
\widetilde{\boldsymbol{\nu}} = \left( \begin{array}{rrr}
\widetilde{\nu}_x & 0 & 0\\
0 & \widetilde{\nu}_y & 0 \\
0 & 0 & \widetilde{\nu}_z  \\
\end{array}\right),
\label{eq:nu_tensor_3d}
\end{equation}
where $\widetilde{\nu}_x, \widetilde{\nu}_y, \widetilde{\nu}_z$ are considered to be constant in the case of a global weighting factor.
We assume the measurement data to be sorted in ascending order such that $(H^\star_{{d},m},B^\star_{{d},m}) < (H^\star_{{d},m+1},B^\star_{{d},m+1}),\,m=1,\dots,M-1$, holds element-wise, i.e. $H^\star_{{d},m} < H^\star_{{d},m+1}$ and $B^\star_{{d},m} < B^\star_{{d},m+1}$. 
Assuming that the data are noiseless and sampled equidistantly, we can apply the centered finite differences scheme to obtain $\boldsymbol{\nu}(\mathbf{B})$, such that
\begin{equation}
\nu_{d}(B_{d}) = \frac{{\mathrm{d}} H_{d}}{{\mathrm{d}} B_{d}} \approx \nu_{{d},m}(B_{{d},m}^\star) = \frac{H_{d}^\star(B^\star_{{d},m}+\Delta B_{d}^\star) - H_{d}^\star(B^\star_{{d},m}-\Delta B_{d}^\star)}{2 \Delta B_{d}^\star}, \quad \text{for~}m=2\dots M-1,
\label{eq:centered_diff}
\end{equation}
with $\Delta B_{d}^\star = B^\star_{{d},m+1} - B^\star_{{d},m} = \text{const.}, \,\forall \, m=1\dots M-1$.
To obtain a weighting factor on the boundaries, i.e. $m=1$, respectively $m=M$, forward or backward differences can be employed. In the case of non-equidistant data, we can apply standard forward differences, thus obtaining
\begin{equation}
\nu_{{d},m}(B_{{d},m}^\star) =  \frac{H_{d}^\star(B^\star_{{d},m}+\Delta B_{d}^\star)- H_{d}^\star(B^\star_{{d},m})}{\Delta B_{d}^\star}, \quad \text{with~}\Delta B_{d}^\star = B^\star_{{d},m+1} - B^\star_{{d},m}, \quad \text{for~}m=1\dots M-1.
\label{eq:forward_diff}
\end{equation}
The stiffness matrix \eqref{eq:stiffnes_matrix}, the \gls{rhs} \eqref{eq:rhs_1_discrete}, and the distance function \eqref{eq:distance_discrete} are then initialized with the mean over the discrete reluctivity curve. 
Thus, when employing \eqref{eq:forward_diff}, the elements of the reluctivity tensor are given as 
\begin{equation}
\widetilde{\nu}_{d} = \mean[\boldsymbol{\nu}_{d}], \quad \mathrm{with~} \boldsymbol{\nu}_{d} = [\nu_{{d},1},\dots,\nu_{{d},M-1}]^\top
\end{equation}
where $\mean[X] = \frac{1}{N} \sum_{n=1}^{N} x_n$. Averaging over all differential reluctivities is of course only reasonable if the data points are almost equidistant. Particularly for data sets with clusters, a weighted average is necessary. Note that, when using a global weighting factor, we assign each element the same computational constant, i.e. one constant represents the entire $BH$-curve. The algorithm employing one global weighting factor is shown in Algorithm~\ref{alg:local_dd}, with the exception of the gray shaded part which is reserved for the local weighting factors assignment, discussed next.

\subsection{Local weighting factors}
\label{subsec:local_mat}
\agcomment{
In the special case of a linear material relation and given a data set $\mathcal{D}$ that represents the closed-form material law with an infinite number of data points, it has been shown that the solution of the data-driven solver is independent of the norm \eqref{eq:distance} \cite{conti2018data}.	 
Contrarily, in the case of finite data, properly chosen weighting factors result in improved convergence rates, as we will show numerically in the following.
Particularly in cases where the states in the measurement set $\mathcal{D}$ depart from the Maxwell-conforming set $\mathcal{M}$, the global weighting factor introduced in Section~\ref{subsec:global_mat} faces major difficulties to produce an accurate solution.
This is especially relevant for strongly nonlinear material responses, where the nonlinearity can lead to certain regions of the material curve being densely sampled, thus the corresponding data set can be characterized as ``well behaved'', while other regions are represented only by sparse and irregular data points \cite{arpaia2020}.
In the following we analyze the behavior of the data-driven solver exactly for such cases.}

\agcomment{
Let us assume that a state $\zeta^\times \in \mathcal{D}$ has been identified as the closest to conform with  Maxwell's equations. 
In the next iterative step, $\zeta^\times$ is incorporated in the \glspl{rhs} of \eqref{eq:rhss}. 
Solving the linear systems \eqref{eq:weak_form_with_zero_kernel} leads to an updated field solution \eqref{eq:update_field} that conforms with Maxwell's equations, i.e. $\zeta^\circ \in \mathcal{M}$. 
We further assume that the chosen material state $\zeta^\times \in \mathcal{D}$ is distant from states in $\mathcal{M}$. 
We can then observe that the updated Maxwell-conforming field solution is also distant from the states given by the material data set, therefore, further iterations do not lead to an improvement. 
In this case, the data set $\mathcal{D}$ does not provide a state that is sufficiently compatible with the governing equations and the global weighting factor in the system matrices \eqref{eq:discrete_problem_definition} project the updated field solution on a fictitious material given by the global weighting factor.}

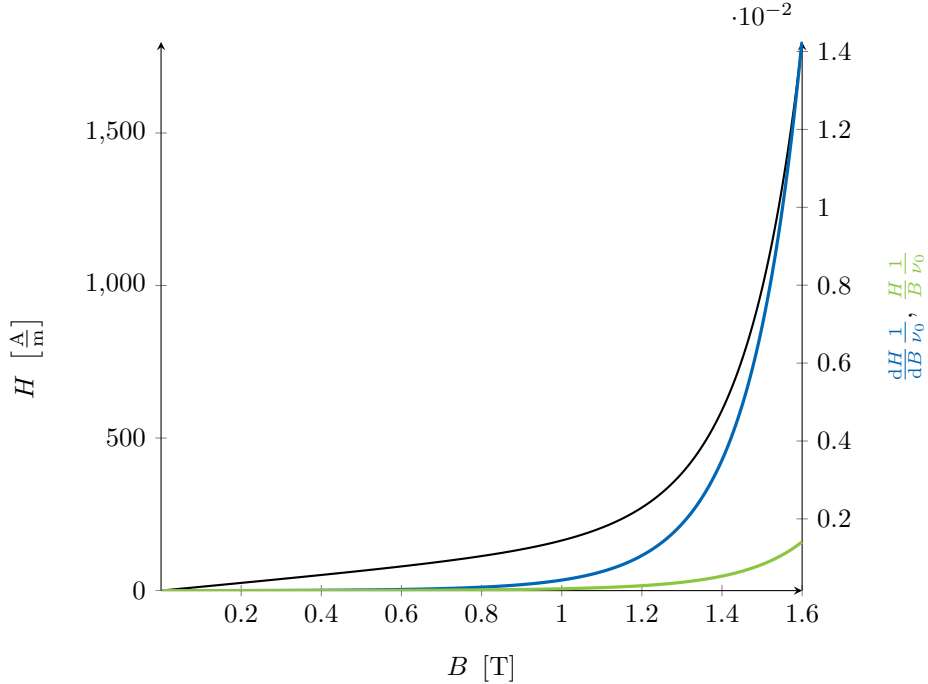
\begin{figure}[t!]
	\begin{center}
	\begin{tikzpicture}
	\centering
	\pgfplotsset{
		scale only axis,
	}	
	\begin{axis}[axis lines=middle,xmin=0, ymin=0,axis y line*=left,xlabel={$B\,\,\left[\mathrm{T}\right]$},ylabel={$H\,\,\left[\frac{\mathrm{A}}{\mathrm{m}}\right]$},x label style={at={(axis description cs:0.5,-0.1)},anchor=north},y label style={at={(axis description cs:-0.25,0.425)},anchor=north, rotate=90}]
		\addplot[thick, black] table[x index=1, y index=0]{figures/HBmu.txt};
	\end{axis}
	\pgfplotsset{
		every axis label/.append style ={black},
		every tick label/.append style={black}  
	}
	\begin{axis}[axis lines=middle,xmin=0,axis y line*=right,axis x line=none,ylabel={\textcolor{TUDa-1b}{$\frac{\mathrm{d}H}{\mathrm{d}B}\frac{1}{\nu_0}$}, \textcolor{TUDa-4b}{$\frac{H}{B}\frac{1}{\nu_0}$}} , y label style={at={(axis description cs:1.12,0.5)},anchor=north, rotate=90}]
		
		\addplot[very thick, TUDa-1b] table[x index=1, y expr=(1/\thisrowno{2})]{figures/HBmu.txt};
		\addplot[very thick, TUDa-4b] table[x index=1, y expr=(\thisrowno{0}/\thisrowno{1}*(4*pi*1e-7))]{figures/HBmu.txt};
	\end{axis}	
\end{tikzpicture}
		\caption{Nonlinear $BH$-curve, relative differential reluctivity $\frac{\mathrm{d}H}{\mathrm{d}B}\frac{1}{\nu_0}$ and relative chord reluctivity $\frac{H}{B}\frac{1}{\nu_0}$.}
		\label{fig:BHmu}
	\end{center}
\end{figure}

\agcomment{We consider here the case of soft magnetic materials, the $BH$-curves of which feature a steep linear part, followed by a sharp transition into saturation. 
We expect that similar material behavior can be observed for material properties in other domains of physics as well.
For ferromagnetic materials, the relative reluctivity in the linear part takes values typically in the region of $\approx 10^{-3}$, whereas in the saturation part its value lies in the region of $\approx 10^{-2}$, until it further increases to $1$ at full saturation.
It is therefore obvious that an averaged differential reluctivity is not capable to cover the $BH$-curve in its entirety, as also illustrated in Figure~\ref{fig:BHmu}, due to the fact that a global weighting factor covers only one operation point in the $BH$-curve.}

As a remedy to this problem, we propose to allocate a local weighting factor per finite element.
We denote the heterogeneous weighting tensor with $\widehat{\boldsymbol{\mu}}$, respectively $\widehat{\boldsymbol{\nu}}$ for reluctivity. 
Since we have no a priori knowledge about the working points of each element, the assignment of local weighting factors must be performed adaptively during the data-driven iteration.
The factors are initialized in the same manner as described in Section \ref{subsec:global_mat}. 
During the iterative process, we monitor the convergence of the distance between the states $\zeta^\circ \in \mathcal{M}$ fulfilling Maxwell's equations and the states $\zeta^\times=(H^\times,B^\times) \in \mathcal{D}$ in the measurement data set, when minimizing the energy mismatch 
\begin{equation}
\epsilon_{\text{em},d} = \sum_{e=1}^{N_\text{elem}} V_e\left[ \frac{1}{2}\widetilde{\mu}_{d,e}\left(H^\circ_{d,e}-H_{d,e}^\times\right)^2 + \frac{1}{2}\widetilde{\nu}_{d,e}\left(B^\circ_{d,e}-B_{d,e}^\times\right)^2\right],
\label{eq:er_energy_mismatch}
\end{equation}
for each dimension ${d}\in \{x,y,z\}$, where $V_e$ refers to the volume of element $e$. 

A stagnation of the energy mismatch \eqref{eq:er_energy_mismatch} during iteration can be attributed to two reasons. 
First, for elements whose $BH$-operating point is matched to the global weighting factor, the states $\zeta^\circ$ are likely to be close to some measurement point in $\mathcal{D}$. 
Second, for elements whose $BH$-operating point is far off the global weighting factor, the states $\zeta^\circ$ are most probably distant to the measurement points and further iterations cannot contribute to an improvement, which is illustrated for an example case in Figure~\ref{fig:global_distance}.
\agcomment{We define the stagnation indicator $\varsigma$ as}
\begin{equation}
\agcomment{\varsigma = \max_{d \in \{x,y,z\}}\left\{\frac{\left|\epsilon^{i}_{\text{em},d} - \epsilon^{i+1}_{\text{em},d}\right|}{\epsilon^{i}_{\text{em},d}}\right\},}
\end{equation}
\agcomment{where the index $i$ refers to the data-driven iterations.}
\agcomment{Stagnation is detected if the stagnation indicator $\varsigma$ becomes smaller than a user defined bound $\varsigma_{\mathrm{bnd}}$}. 
In that case, a switch from global to local weighting factors is triggered, where the local reluctivity is estimated using the local material solution, i.e. with the states $(\mathbf{H}^\times,\mathbf{B}^\times)$ selected by the energy mismatch minimization. 
Note that it is also possible to switch from global to local weighting factors after a user-defined number of iterations. 
Some iterations, e.g. $<10$, with a globally averaged weighting factor are actually desirable, since the states $\zeta^\times$ are randomly initialized on $\mathcal{D}$, consequently, an early local treatment could estimate factors that are not optimal.
The local weighting factors are given by
\begin{align}
\widehat{\nu}_{{d}}(\vec{x}) &= \sum_{e=1}^{N_\mathrm{elem}} \nu_{{d}}\left(B^\times_{{d,e}}\right) \mathbbm{1}_e(\vec{x}) \quad \mathrm{with~} {d}\in \{x,y,z\}, \label{eq:local_nu}\\
\mathbbm{1}_e(\vec{x}) &= \left\{\begin{array}{ll} 1, & \vec{x}\in T_e \\
0, & \vec{x}\not\in T_e\end{array}\right. ,
\end{align}
where $T_e$ refers to the $e$-th element of the triangulation of $\Omega$.
Keeping these considerations in mind, we adapt the data-driven algorithm as shown in Algorithm~\ref{alg:local_dd}.
Now, each element has a specific local weighting factor $\widehat{\nu}_{d}(\vec{x}_e)$, estimated with \eqref{eq:centered_diff} or \eqref{eq:forward_diff}. 
This process seems similar to the material constant assigned with a standard Newton solver. 
However, we point out that, in contrast to a Newton solver, we do not explicitly model the material and, more importantly, we do not rely on a physical representation of the material, \agcomment{except for the bounds given in \eqref{eq:mu_constraints}}.
Since the weighting factors are now chosen optimally with respect to the current local field solution for each element, we can expect an improvement in the convergence of the data-driven solver.

\begin{center}
	\begin{algorithm}[htb]
	\begin{algorithmic}
		\State \textbf{initialize} $(\mathbf{H}^\times, \mathbf{B}^\times)$ randomly on $\mathcal{D}$ 
		\State \textbf{estimate} $\widetilde{\boldsymbol{\nu}}$ on $\mathcal{D}$ 
		\While{convergence not reached}
			\vspace{0.5em}
			\State Solve variational problem \eqref{eq:discrete_problem_definition} for $\mathbf{a}$ and $\boldsymbol{\eta}$
			\vspace{0.25em}
			\State Update $\vec{B}_h =\mathrm{curl}\vec{a}_h$ 
			\State \phantom{Update }$\vec{H}_h = \vec{H}_h^\times +\widetilde{\boldsymbol{\nu}}\mathrm{curl} \vec{\eta}_h$ 
			\vspace{0.25em}
			\State Find $(\Hx, \Bx)$ in $\mathcal{D}$ adjacent to $(\Ho, \Bo)$ with distance function \eqref{eq:distance} 
			\vspace{0.5em}
			\If{\tikzmark{starta} stagnation detected}
				\State assign local reluctivity $\widehat{\boldsymbol{\nu}}$ to elements \eqref{eq:local_nu}
				\State assemble stiffness matrix $\mathbf{K}_{\hat{\nu}}$ 
				\State update distance function \eqref{eq:distance} with local reluctivity $\widehat{\boldsymbol{\nu}}$\tikzmark{enda}
			\EndIf
		\EndWhile
		\tikz[remember picture,overlay,pin distance=0cm]
		{\draw[line width=1pt,draw=black,fill=black,rectangle,rounded corners,fill opacity=0.1]
			([xshift=-15pt,yshift=10pt]$ (pic cs:starta)$ ) rectangle ([xshift=5pt,yshift=-16pt]$ (pic cs:enda)$ );
		\draw [decorate,decoration={brace,amplitude=10pt, mirror}]
		( $ (pic cs:enda) + (1ex,-3.7ex) $ ) --  ( $ (pic cs:enda) + (1ex,11.2ex) $ )  node  [align=left,xshift=2ex,black,right,pos=0.5] 
		{\emph{only for local weighting} \\ \emph{factor assignment}};
		}
	\end{algorithmic}
		\caption{Data-driven magnetostatic field solver. The gray part shows the extension according to the local weighting factors.}  
		\label{alg:local_dd}	
\end{algorithm}
\end{center}

\subsection{Numerical experiments}
\label{subsec:noisefree_numerical_experiments}
\begin{figure}[htb]
	\begin{subfigure}[t]{0.47\textwidth}
	\centering
	\includegraphics[width=0.6\textwidth]{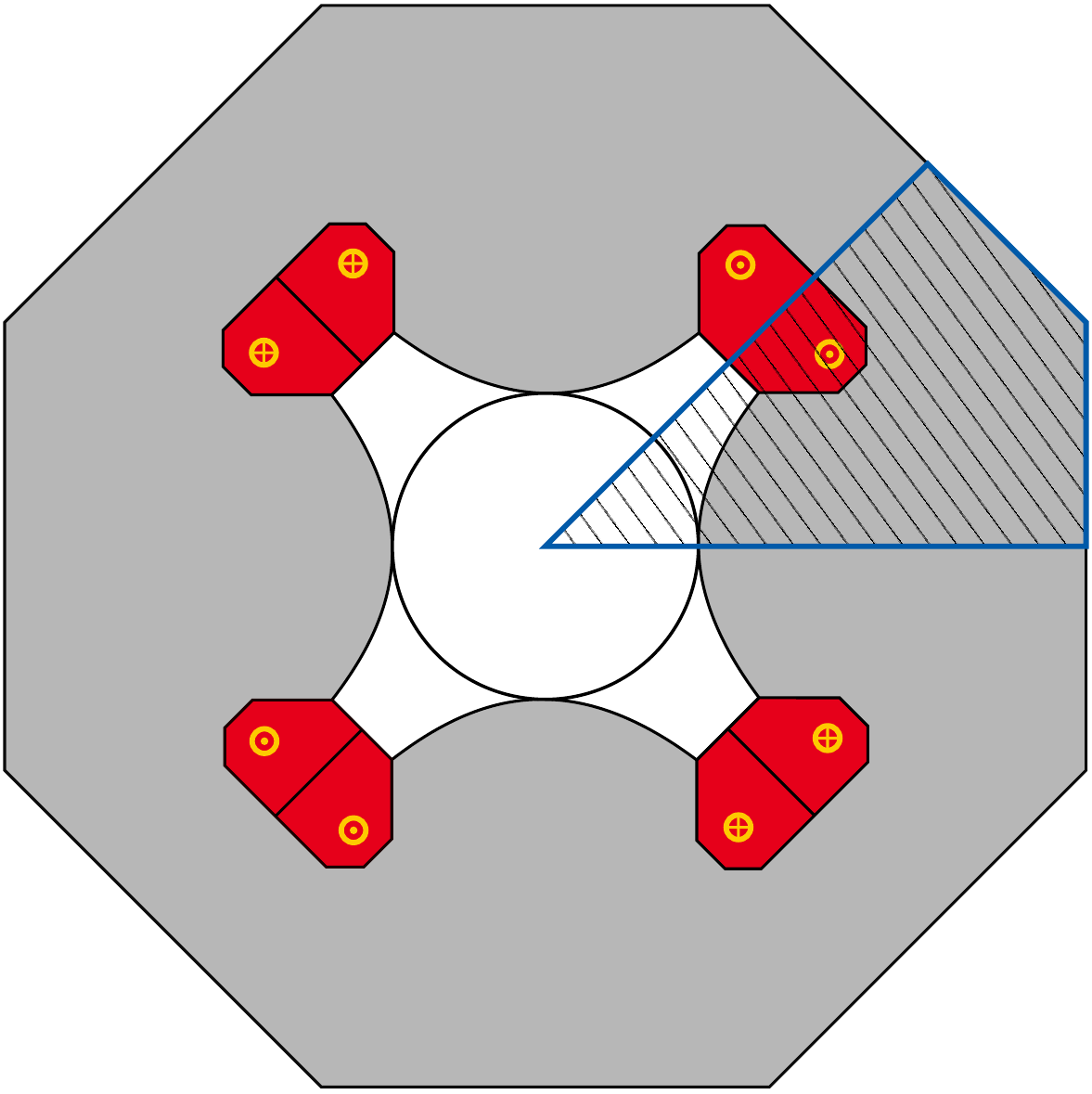}
	\caption{}
	\label{fig:quad_geometry}
	\end{subfigure}
	\hfill
	\begin{subfigure}[t]{0.47\textwidth}
	\centering
	\begin{tikzpicture}[x=1cm,y=1cm]
	\def\a{6.5}
	\node at (2.3*\a,2.3*\a)       {\includegraphics[width=6.5cm]{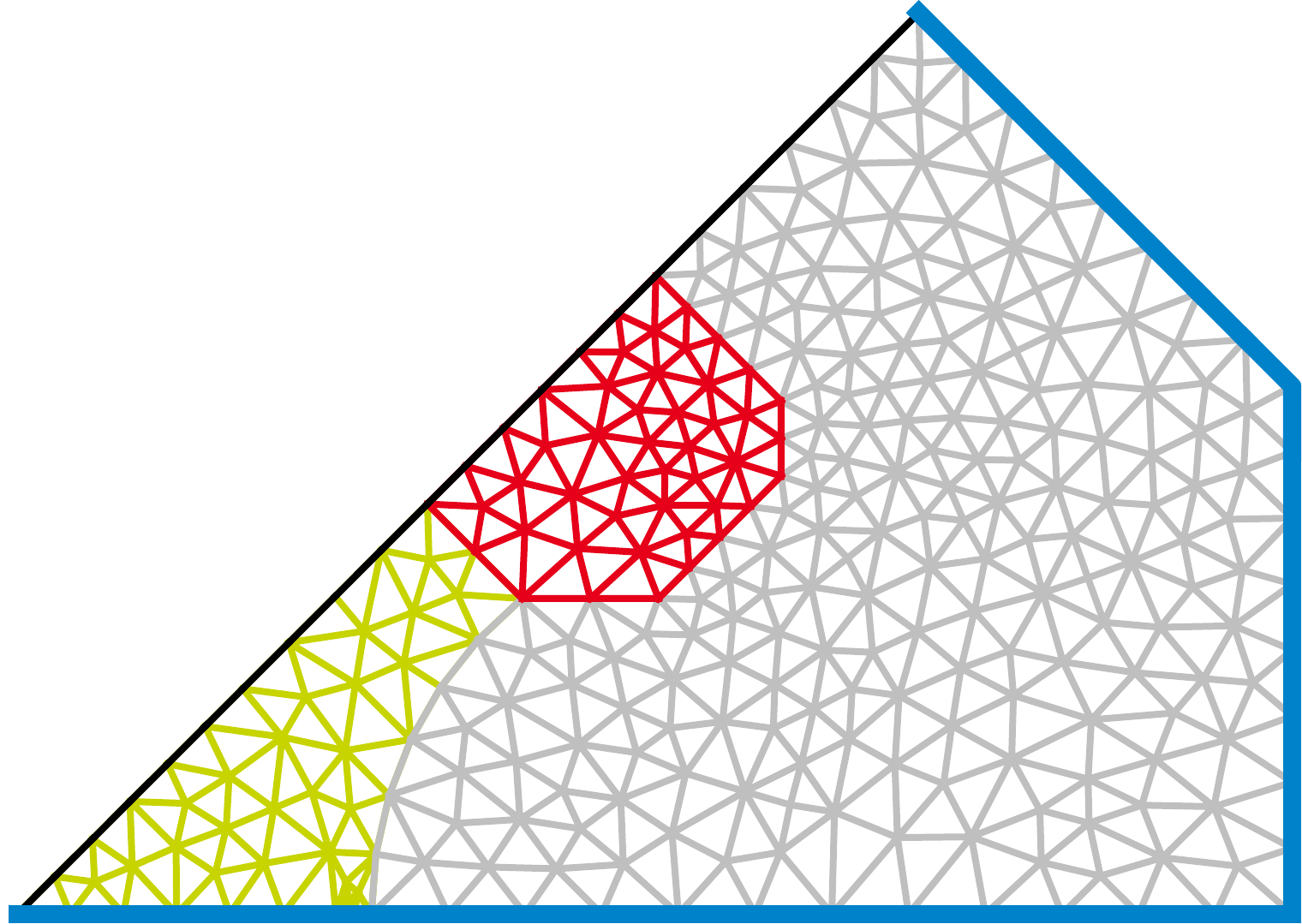}};
	\node at (2.8*\a  , 2.55*\a) {$B_n|_{\Gamma_\text{D}}=0$};
	\node at (1.95*\a   , 2.3*\a) {$H_t|_{\Gamma_\text{N}}=0$};
	\end{tikzpicture}	
	\caption{}
	\label{fig:quad_mesh}
	\end{subfigure}	
	\caption{(a) Cross-section of a quadrupole magnet. The gray area is iron, the red areas are the coils, and the white area is vacuum. The shaded polygon with the blue boundary shows the computational domain. (b) Computational domain and mesh according to one eighth of the quadrupole magnet. On the blue boundary, homogeneous Dirichlet boundary conditions are imposed, whereas on the black boundary homogeneous Neumann conditions are imposed.}
	\label{fig:quad}		
\end{figure} 

For our numerical investigations, we employ the model of a quadrupole magnet. 
Such magnets are used in accelerator facilities to focus particle beams.
Due to translational invariance along the $z$-direction, the magnet can be modeled in 2D without significant compromises in terms of modeling accuracy.
The 2D magnet model is depicted in Figure~\ref{fig:quad_geometry} and consists of three different domains, namely, the iron yoke (gray), the coils (red), and vacuum (white).  
Exploiting the symmetry of the model and assuming that the permeability of the yoke is sufficiently high, such that the magnetic flux leaving the yoke can be neglected, it is sufficient to consider only one eighth of the quadrupole's geometry, shown in Figures~\ref{fig:quad_geometry} (marked area) and \ref{fig:quad_mesh}.
The iron yoke has an anisotropic permeability, which is linear in the $y$-direction \agcomment{with a relative permeability of $\mu_\mathrm{r}=300$, and nonlinear in the $x$-direction following the Brauer model \cite{brauer1975}}
\begin{equation}
\agcomment{	H_x\left(B_x\right) = \nu\left(B_x\right)B_x = \left(k_1 e^{k_2B_x^2}+k_3\right)B_x},
\label{eq:brauer_model}
\end{equation}
\agcomment{with $k_1=\SI{6}{\meter\per\henry}, k_2=\SI{2}{\per\square\tesla}, k_3=\SI{120}{\meter\per\henry}$.}
The material responses of the coils and vacuum are considered to be exactly known, thus, those regions are treated as described in Section~\ref{sec:sec_exactly_known_material}.
The magnetic material is fully treated by the data-driven solver as discussed in Section~\ref{subsec:discrete_dd}.
The permeability of the yoke follows the behavior of a soft magnetic material in the $x$-direction.
Due to their small hysteresis, soft magnetic materials are widely used in inductors, electrical machines, and other devices to reduce losses. 
Moreover, soft magnetic materials feature strongly nonlinear magnetization curves. 

\begin{figure}[h!]
	\begin{subfigure}[t]{0.48\textwidth}
		\includegraphics[width=0.95\textwidth]{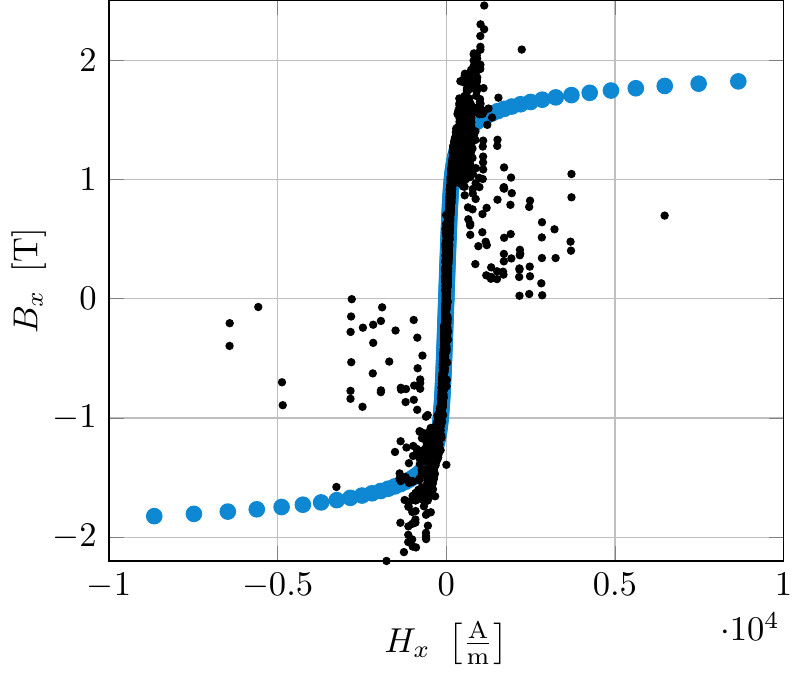} 
		\caption{}
		\label{fig:global_distance}
	\end{subfigure}
	\hfill
	\begin{subfigure}[t]{0.48\textwidth}
		\includegraphics[width=0.95\textwidth]{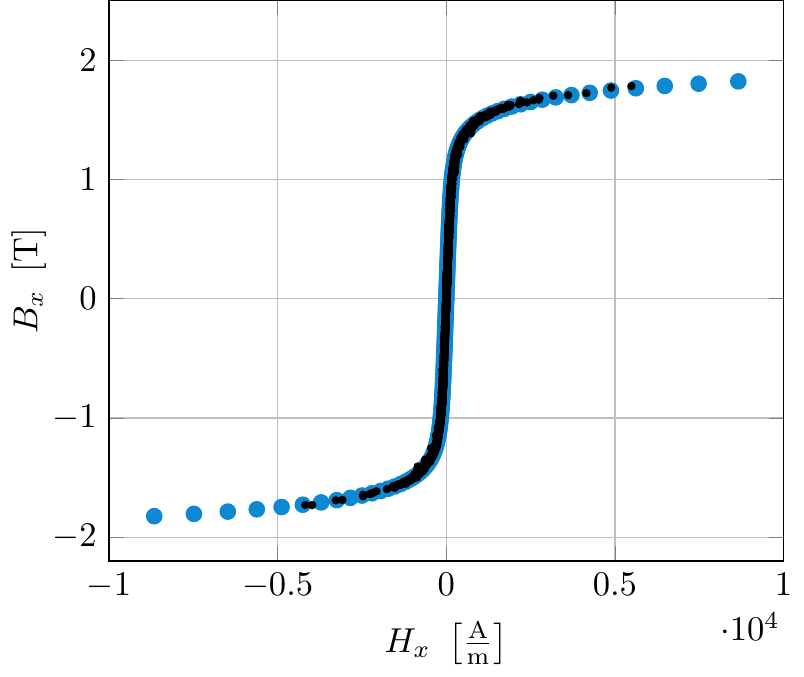} 
		\caption{}
		\label{fig:local_distance}
	\end{subfigure}
	\caption{$(H_x,B_x)$-pairs corresponding to measurement data (\textcolor{TUDa-2b}{blue points}) and to the data-driven field solution after solver convergence  \textcolor{black}{(black points)}. Both data-driven simulations have been carried out with $N=10^2$ data points. (a) Global weighting factor $\widetilde{\boldsymbol{\nu}}$. (b) Local weighting factors $\widehat{\boldsymbol{\nu}}$.}
	\label{fig:distance}
\end{figure}

Both data-driven solvers, i.e. respectively utilizing global or local weighting factors, are employed to provide a data-driven solution of the quadrupole's magnetic field.
The data points in $\mathcal{D}$ are generated by sampling the material curves in $x$- and $y$-direction equidistantly. \agcomment{Due to the linear material behavior in $y$-direction, the sampling points are spread equidistantly along the material curve. However, this is not true for the nonlinear relation of the $x$-direction, see Figure~\ref{fig:distance}. Here, $B$ is sampled equidistantly, while $H(B)$ follows \eqref{eq:brauer_model}, hereby loosing the equidistance property along the curve. This leads to regions that are data-rich, while the saturation part remains data-starved.} 
As illustrated next, the original data-driven algorithm \cite{kirchdoerfer2016data}, equivalently, Algorithm \ref{alg:local_dd} in this paper, has difficulties in providing accurate solutions, due to the fact that the global weighting factor $\widetilde{\boldsymbol{\nu}}$ which applies to the entire domain $\Omega$ must fit not only to the linear part of the $BH$-curve, but also to the saturation part as well. 
On the contrary, the data-driven solver with local weighting factors is able to resolve the nonlinearity of the material law, providing significant efficiency gains.

To verify the results obtained with either data-driven field solver, we introduce three error metrics. The first error quantifies the energy mismatch between the most recently computed states $\zeta^\circ \in \mathcal{M}$ fulfilling the governing equations and the reference solution, and is defined as
\begin{equation}
\epsilon_\text{em} = \left(\frac{\langle \Ho - \mathbf{H}_\text{ref} , \Ho - \mathbf{H}_\text{ref} \rangle_{\mathbf{D}_{{\boldsymbol{\nu}}_\mathrm{ref}}} + \langle \Bo - \mathbf{B}_\text{ref}, \Bo - \mathbf{B}_\text{ref} \rangle_{\mathbf{D}_{{\boldsymbol{\mu}}_\mathrm{ref}}}}{\langle \mathbf{H}_\text{ref}, \mathbf{H}_\text{ref} \rangle_{\mathbf{D}_{{\boldsymbol{\nu}}_\mathrm{ref}}} + \langle \mathbf{B}_\text{ref}, \mathbf{B}_\text{ref} \rangle_{\mathbf{D}_{{\boldsymbol{\mu}}_\mathrm{ref}}}}\right)^\frac{1}{2} 
\label{eq:err_em}
\end{equation}
The other two metrics are the relative field errors
\begin{align}
\epsilon_{H} &=  \left(\frac{\langle \Ho - \mathbf{H}_\text{ref} , \Ho - \mathbf{H}_\text{ref} \rangle_{\mathbf{D}_{{\boldsymbol{\nu}}_\mathrm{ref}}}}{\langle  \mathbf{H}_\text{ref}, \mathbf{H}_\text{ref} \rangle_{\mathbf{D}_{{\boldsymbol{\nu}}_\mathrm{ref}}} }\right)^\frac{1}{2}, \label{eq:err_rms_H} \\
\epsilon_{B} &=  \left(\frac{\langle \Bo - \mathbf{B}_\text{ref} , \Bo - \mathbf{B}_\text{ref} \rangle_{\mathbf{D}_{{\boldsymbol{\mu}}_\mathrm{ref}}}}{\langle \mathbf{B}_\text{ref}, \mathbf{B}_\text{ref} \rangle_{\mathbf{D}_{{\boldsymbol{\mu}}_\mathrm{ref}}}}\right)^\frac{1}{2},
\label{eq:err_rms_B}
\end{align}
with respect to the magnetic field strength $\vec{H}$ and the magnetic flux density $\vec{B}$.
Note that $\mathbf{D}_{{\boldsymbol{\mu}}_\text{ref}}$ and $\mathbf{D}_{{\boldsymbol{\nu}}_\text{ref}}$ refer to the permeability and reluctivity coefficients obtained from the reference solution, which are computed with a standard Newton solver. 

To monitor the convergence of the errors defined in \eqref{eq:err_em}, \eqref{eq:err_rms_H}, and \eqref{eq:err_rms_B}, the data-driven solvers are employed with measurement data sets $\mathcal{D}$ of increasing size. \agcomment{For all employed data sets, the data-driven solver switches from global to local weighting factors based on a user-defined stagnation bound $\varsigma_{\mathrm{bnd}}=10^{-2}$. 
Switching typically occurs at approximately $20$ iterations. 
For better comparability of the results presented next, the switching point is fixed to $20$ iterations in all simulations. 
However, we note that, after a small number of iterations with global weighting factors, typically $4-5$, the switching point has a negligible influence on the final solution, at least for the examples considered in this work.}
\agcomment{Moreover, we note that a convergence study with respect to the mesh size is omitted in this work, since numerical tests and theoretical considerations have already concluded that no notable difference in the data-driven solution is to be expected \cite{kirchdoerfer2016data}. We can also confirm that conclusion, based on own numerical tests with the quadrupole magnet model.}

\begin{figure}[t!]
	\begin{subfigure}[t]{0.48\textwidth}
		\begin{tikzpicture}
		\centering
		\begin{axis}[xmode=log,ymode=log,xlabel={Number of data points},ylabel={Energy mismatch $\epsilon_\text{em}$}, legend pos=south west]
		\addplot[mark=*, mark options={solid}, thick, TUDa-1b, dashed] table[x index=0, y index=1]{figures/quadrupole/global/quad_global_dist.txt};
		\addlegendentry{global}
		\addplot[mark=*, thick, TUDa-1b] table[x index=0, y index=1]{figures/quadrupole/local/quad_local_dist.txt};
		\addlegendentry{local}
		
		\draw[black] (axis cs:1e+2,1e-4) -- (axis cs:1e+3,1e-4);
		\draw[black] (axis cs:1e+2,1e-4) -- (axis cs:1e+2,1e-3);
		\draw[black] (axis cs:1e+2,1e-3) -- (axis cs:1e+3,1e-4);
		\node at (axis cs:4e+2, 5e-5) {$\mathcal{O}(N^{-1})$};
		
		\end{axis}
		\end{tikzpicture}
		\caption{}
		\label{fig:quad_conv_distance}
	\end{subfigure}
	\hfill
	\begin{subfigure}[t]{0.48\textwidth}
		\begin{tikzpicture}
		\centering
		\begin{axis}[legend columns=2, legend style={/tikz/column 2/.style={column sep=5pt}},xmode=log,ymode=log,xlabel={Number of data points},ylabel={Relative errors $\epsilon_{H}, \epsilon_{B}$}, legend pos=south west]
		\addlegendimage{empty legend}
		\addlegendentry{\hspace{-.6cm}\textbf{global}}
		\addlegendimage{empty legend}
		\addlegendentry{\hspace{-.6cm}\textbf{local}}
		\addplot[mark=*, mark options={solid}, thick, TUDa-7b, dashed] table[x index=0, y index=1]{figures/quadrupole/global/quad_global_Hrel.txt};, legend style={at={(1.0,1.0)},anchor=north east}
		\addlegendentry{$\epsilon_{H}$} 
		\addplot[mark=*, thick, TUDa-7b] table[x index=0, y index=1]{figures/quadrupole/local/quad_local_Hrel.txt};
		\addlegendentry{$\epsilon_{H}$} 
		\addplot[mark=*, mark options={solid}, thick, TUDa-10b, dashed] table[x index=0, y index=1]{figures/quadrupole/global/quad_global_Brel.txt};
		\addlegendentry{$\epsilon_{B}$} 
		\addplot[mark=*, thick, TUDa-10b] table[x index=0, y index=1]{figures/quadrupole/local/quad_local_Brel.txt};
		\addlegendentry{$\epsilon_{B}$} 
		\end{axis}
		\end{tikzpicture}
		\caption{}
		\label{fig:quad_rms}
	\end{subfigure}	
	\caption{Convergence of data-driven solutions with respect to data set size. The solid lines refer to the data-driven solver utilizing local weighting factors, while the dashed lines refer to the standard data-driven solver based on a global weighting factor. (a) Convergence of the energy mismatch for increasing data set size. (b) Convergence of the relative errors in the $\vec{H}$- and $\vec{B}$-field for increasing data set size.}
	\label{fig:quad_noisfree_conv}
\end{figure}
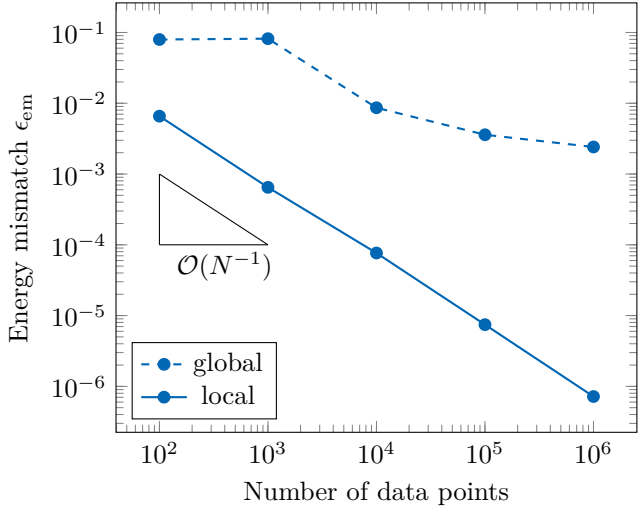
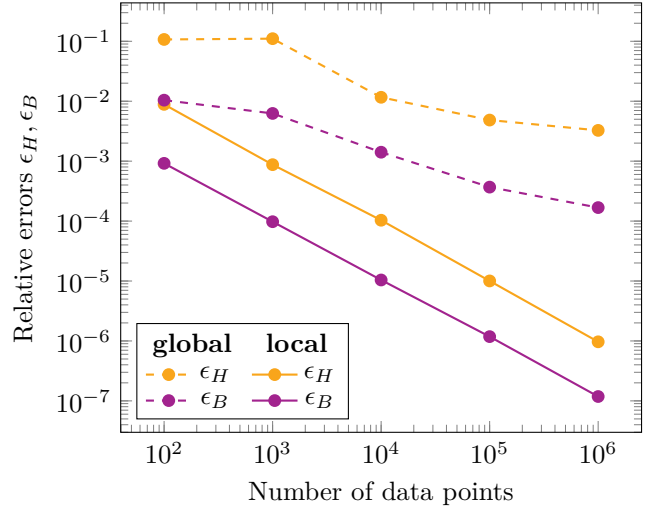

The convergence behavior of both data-driven solvers with respect to all three errors is shown in Figures~\ref{fig:quad_conv_distance} and \ref{fig:quad_rms}.
It can easily be observed that, for the same data set, the data-driven solver utilizing local weighting factors is orders of magnitude more accurate than the original, with respect to all three error metrics.
In particular, local weighting factors lead to a linear convergence rate.
Moreover, the accuracy gain due to the use of local weighting factors becomes even more pronounced as the size of the data set increases.

\begin{figure}[t!]
	\begin{subfigure}[t]{0.48\textwidth}
		\includegraphics[width=0.95\textwidth]{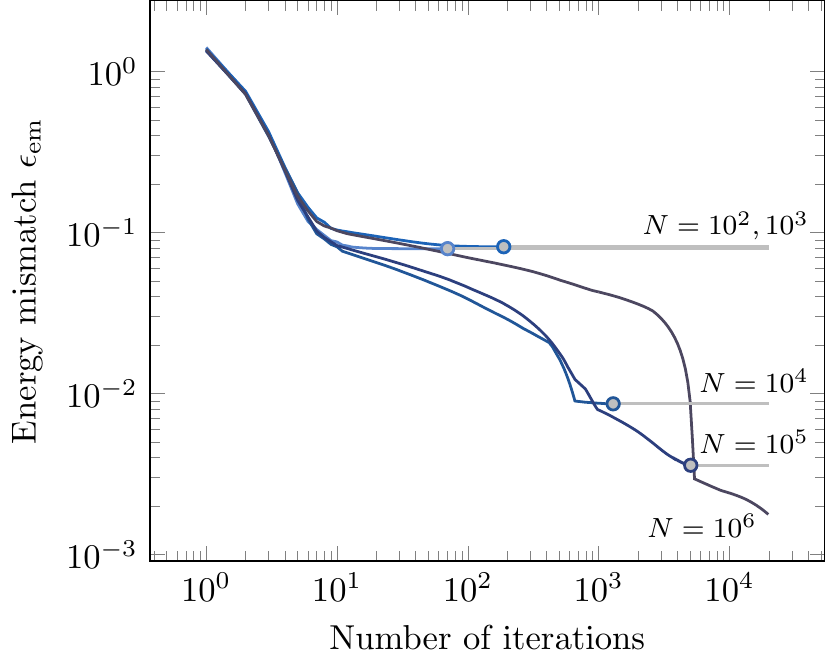} 
		\caption{}
		\label{fig:global_quad_conv_distance_over_iter}
	\end{subfigure}
	\hfill
	\begin{subfigure}[t]{0.48\textwidth}
		\includegraphics[width=0.95\textwidth]{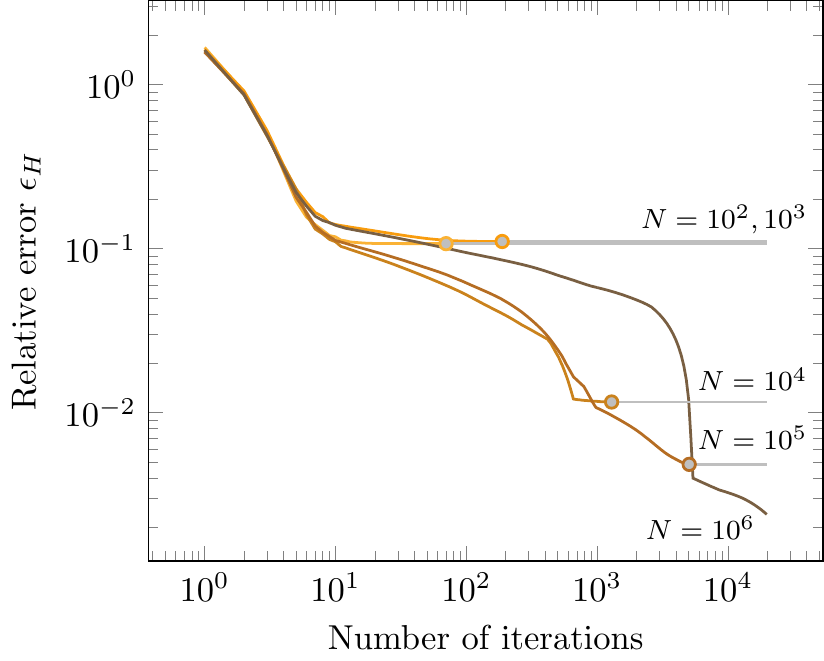} 
		\caption{}
		\label{fig:global_quad_rms_over_iter}
	\end{subfigure}		
	\caption{Convergence of the data-driven solution with respect to the number of solver iterations for the case of a global weighting factor. The dots indicate that convergence is reached. (a) Convergence of the energy mismatch for increasing solver iterations over data sets of increasing size $N$. (b) Convergence of the relative error in the $\vec{H}$-field for increasing solver iterations over data sets of increasing size $N$.}
	\label{fig:quad_noisefree_global_iterations}
\end{figure}

\begin{figure}[h!]
	\begin{subfigure}[t]{0.48\textwidth}
		\includegraphics[width=0.95\textwidth]{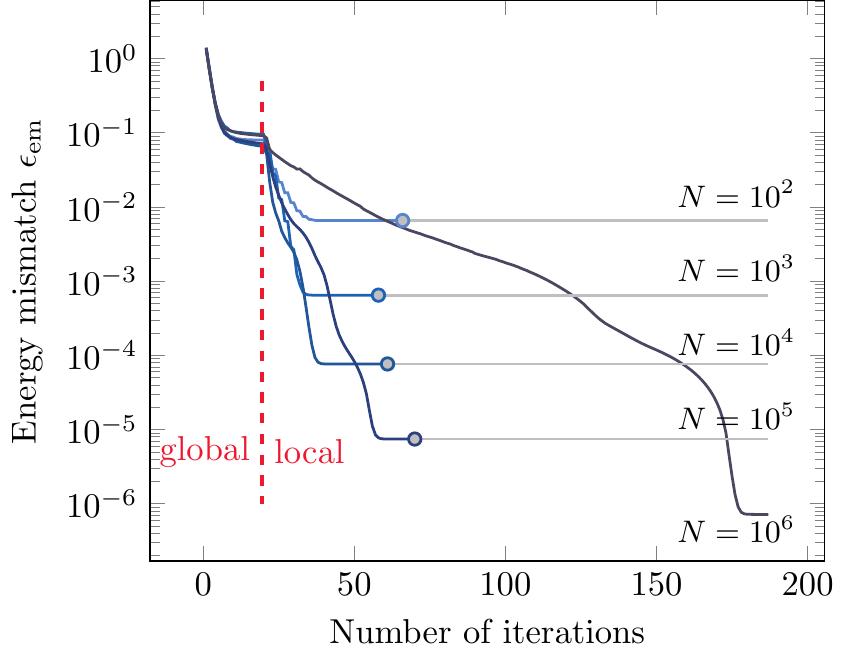} 
		\caption{}
		\label{fig:quad_conv_distance_over_iter}
	\end{subfigure}
	\hfill
	\begin{subfigure}[t]{0.48\textwidth}
		\includegraphics[width=0.95\textwidth]{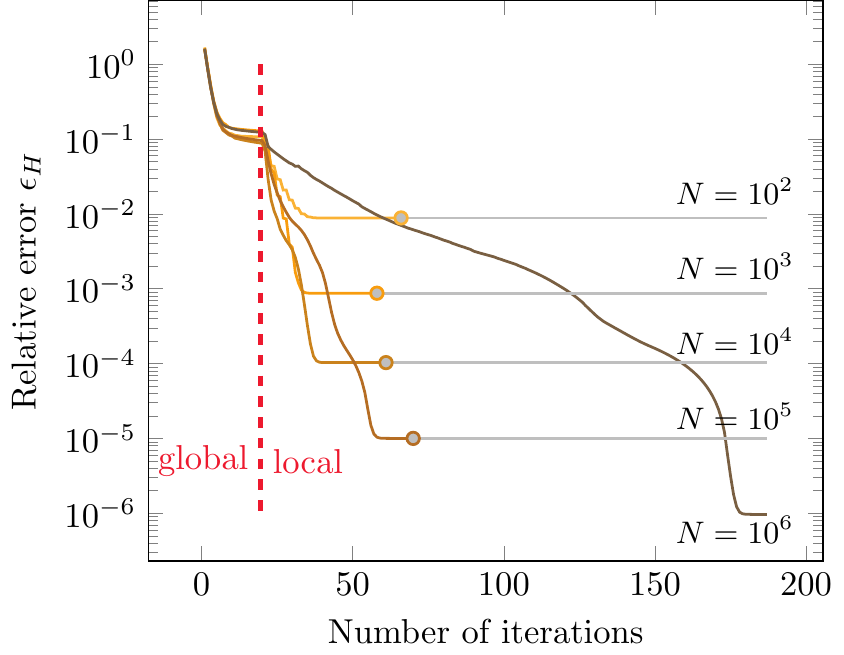} 
		\caption{}
		\label{fig:quad_rms_over_iter}
	\end{subfigure}		
	\caption{Convergence of the data-driven solution with respect to the number of solver iterations for the case of local weighting factors. The dots indicate that convergence is reached. The algorithm switches from global to local material assignment after $20$ iterations. Note that the horizontal scale differs from the one in Figure~\ref{fig:quad_noisefree_global_iterations}. (a) Convergence of the energy mismatch for increasing solver iterations over data sets of increasing size $N$. (b) Convergence of the relative error in the $\vec{H}$-field for increasing solver iterations over data sets of increasing size $N$.}
	\label{fig:quad_noisefree_local_iterations}	
\end{figure}

The efficiency gains due to the local weighting factors become even more obvious if we additionally observe the number of outer-loop iterations performed by each data-driven solver, again considering data sets of increasing size.
The convergence of the standard data-driven solver with a global weighting factor in terms of outer-loop iterations is shown in  Figure~\ref{fig:quad_noisefree_global_iterations}. 
The convergence behavior of the modified solver utilizing local weighting factors is shown in Figure~\ref{fig:quad_noisefree_local_iterations}.
As can be observed, switching from global to local material assignment, as suggested in Algorithm~\ref{alg:local_dd}, leads to a  substantial improvement in terms of computational demand, now quantified by the number of solver iterations. 
This is especially true in the case of large data sets, i.e. with $10^4$ or more points.
 
This improvement in accuracy and efficiency is to be expected due to the fact that the $x$-component of the $BH$-curve of the soft magnetic material comprising the yoke features a strongly nonlinear behavior.
To better illustrate the deficiency of the standard data-driven solver and the advantages of local material assignment, Figure~\ref{fig:distance} shows the $x$-component of the field solution (black points), i.e. the states $(\Ho_x,\Bo_x)\in \mathcal{M}$ identified by the solver, in the iron part of the quadrupole. 
The available material data are depicted as well (blue points).
Both figures refer to the field solution after each algorithm has converged. 
\agcomment{As predicted, due to the missing information in the saturation region, the standard data-driven solver based on a global weighting factor fails to capture the material response, in particular regarding the saturation region, as shown in Figure~\ref{fig:global_distance}.}
On the contrary, the data-driven solver based on local weighting factors yields states that accurately match the material data and the underlying constitutive relation, thus leading to the accuracy and efficiency improvements shown in Figures~\ref{fig:quad_noisfree_conv} and \ref{fig:quad_noisefree_local_iterations}.
\agcomment{Particularly in the region connecting the linear part of the material curve to the saturation part, a sufficient number of data points is necessary to ensure a data set $\mathcal{D}$ that is compatible with the Maxwell equations.}

\section{Data-Driven Magnetostatic Simulation with Noisy Data}
\label{sec:noisy}
In real-world applications, it is unreasonable to assume that the measurement data are free of noise. Measurement noise is introduced by varying ambient conditions, tolerances of the measurement device, or other factors that can influence the quality of the measurement.  
A common assumption is that the added noise follows a Gaussian distribution.
Here we assume independent and identically distributed Gaussian noise with zero mean and variances $\sigma_{H}$, respectively $\sigma_{B}$, added on the measurement data. 
A noisy measurement sample $m$ is then defined as
\begin{equation}
	\left(H_m^\star,B_m^\star\right) = \left(\hat{H}_m^\star + \varepsilon_H,\hat{B}_m^\star + \varepsilon_{B}\right),
\label{eq:noise}
\end{equation}
where $\hat{H}^\star$, $\hat{B}^\star$ are the noiseless measurements and $\varepsilon_H \sim \mathcal{N}(0,\sigma_{H})$, $\varepsilon_B \sim \mathcal{N}(0,\sigma_{B})$ denote the Gaussian noise terms.

The data-driven methods presented and employed in Section \ref{sec:noisefree} are not suitable for the solution of problems with noisy data due to the fact that the distance-minimization approach is highly sensitive to outliers \cite{kirchdoerfer2017data}. 
That is, since all data points are considered of equal importance, the solver could choose an outlying point, i.e. one far away from material data clusters, which however conforms better to a solution of the Maxwell equations.
Nevertheless, data clusters signify that the corresponding data points are more probable to represent the underlying material law.
Clustered data points should then be considered of increased importance compared to outliers in the data set.
It is therefore of relevance to attach a measure of importance to the single data points.
This is done by first performing a clustering analysis on the measurement set $\mathcal{D}$. 
To identify the clusters, a scheme based on maximum entropy estimation  \cite{jaynes1957, kesavan2009} and simulated annealing \cite{kirkpatrick1983} is used \cite{kirchdoerfer2017data}. 
The same idea is used in the present work also, albeit combined with the local weighting factors introduced in Section~\ref{subsec:local_mat}.

\subsection{Global weighting factor}
\label{subsec:noisy_global}
We assume that a local state $\zeta_e \in \mathcal{M}$ fulfilling the magnetostatic equations in an element $e$ of the triangulation of $\Omega$ is available. 
Prior to performing the distance minimization as in Section~\ref{sec:noisefree}, the measurement data are weighted according to their importance. 
A weighting factor $p_{e,m}$ is assigned to each data point $\zeta^\star_m=(H_m^\star,B_m^\star)$, where it holds that
\begin{subequations}
\begin{align}
	\sum_{m=1}^M p_{e,m} &= 1, \label{eq:weights_one} \\
	p_{e,m} &\in [0,1].
	\label{eq:weights}%
\end{align}
\end{subequations}
The weights should be unbiased, while points distant from the local state $\zeta_e$ should have small weights compared to closer ones.
A distribution satisfying these constraints is given by
\begin{subequations}
	\begin{align}
		p_{e,m}(\zeta_e,\beta)&= Z_e(\zeta_e,\beta)^{-1}\mathrm{e}^{-\frac{\beta}{2}\left|\left|\zeta_e-\zeta^\star_m\right|\right|^2_{\tilde{\boldsymbol{\mu}},\tilde{\boldsymbol{\nu}}}}, \label{eq:boltzmann_weight}\\
		Z_e(\zeta_e,\beta) &= \sum_{m=1}^M \mathrm{e}^{-\frac{\beta}{2}\left|\left|\zeta_e-\zeta^\star_m\right|\right|^2_{\tilde{\boldsymbol{\mu}},\tilde{\boldsymbol{\nu}}}}, \label{eq:partition_function}
	\end{align}
\end{subequations}
where the constant $\beta$ controls the influence of data points distant to the local state $\zeta_e$. The distance weighted center $\bar{\zeta}_e^\times$ per element is then given by
\begin{equation}
	\bar{\zeta}_e^\times = \left(\bar{H}_e^\times,\bar{B}_e^\times\right) = \sum_{m=1}^M p_{e,m}(\zeta_e,\beta)\zeta_m^\star. 
\end{equation}
In the context of machine learning, \eqref{eq:boltzmann_weight} is known as the softmax function \cite{rasmussen2006}, whereas in statistical mechanics it is connected to the Boltzmann distribution. 
The normalization constant \eqref{eq:partition_function} is known as the partition function.
Applying this weighting, the data-driven solver selects clusters in the $\beta$-controlled neighborhood of $\zeta_e$, instead of outliers.
Effectively, the weight $p_{e,m}$ represents the probability that the data point $\zeta_m^\star$ is close to the local state $\zeta_e$. 
For instance, if the measurement point $m$ is matched to the local state, the weight for that data point is $p_{e,m}=1$ and all other weights vanish due to \eqref{eq:weights_one}, i.e. $p_{e,j}=0, \,\forall j=1,\cdots,M\,\land m\neq j$. 
Note that in the case of $\beta \rightarrow \infty$, the classical distance minimization scheme is recovered \cite{kirchdoerfer2017data}.

The extension of the data-driven algorithm is then straightforward. 
Instead of employing the distance minimization function \eqref{eq:distance_discrete}, we seek for the states $\zeta_e \in \mathcal{M}$ that minimize the free energy in phase space, given by
\begin{equation}
	\zeta_{\text{opt},e} \in \argmin_{\zeta_e \in \mathcal{M}} \left\{ P_e(\zeta_e,\beta) \right\}, \quad \text{with~}P_e(\zeta_e,\beta) = -\beta^{-1}\log\left(Z_e(\zeta_e,\beta)\right).
\end{equation}
This procedure is then applied to all local states, i.e. all $(\mathbf{H},\mathbf{B})$-pairs. 
Consequently the total free energy is given by
\begin{equation}
	P(\zeta,\beta) = \sum_{e=1}^{N_\text{elem}}P_e(\zeta_e,\beta) = \sum_{e=1}^{N_\text{elem}} -\beta^{-1} \log\left(Z_e(\zeta_e,\beta)\right).
	\label{eq:total_free_energy}
\end{equation}
 A bottleneck of performing the minimization with respect to the total free energy, is that the function \eqref{eq:total_free_energy} can be strongly non-convex \cite{kirchdoerfer2017data}. 
 Hence, an iterative solver could converge to a local minimum. 
 However, $P(\zeta,\beta)$ is convex for a sufficiently small value of the constant $\beta$.
 The issue can then be resolved by initializing $\beta$ with a suitable value such that $P(\zeta,\beta)$ is convex, and then increasing that value based on simulated annealing \cite{kirkpatrick1983}, until the global minimum is reached \cite{kirchdoerfer2017data}. 

\subsection{Local weighting factors}
\agcomment{For noisy data, local weighting factors not only improve the convergence rate of the data-driven solver, but the solver also produces more robust results, as we will show numerically.}
In contrast to the noiseless case, the data in the noisy case are disordered, making it less trivial to assign the differential reluctivity to a local state $\zeta_e$. 
In the case of ordered data, it is straightforward to apply the tools from differential calculus. 
Otherwise, a beforehand data analysis is essential. 

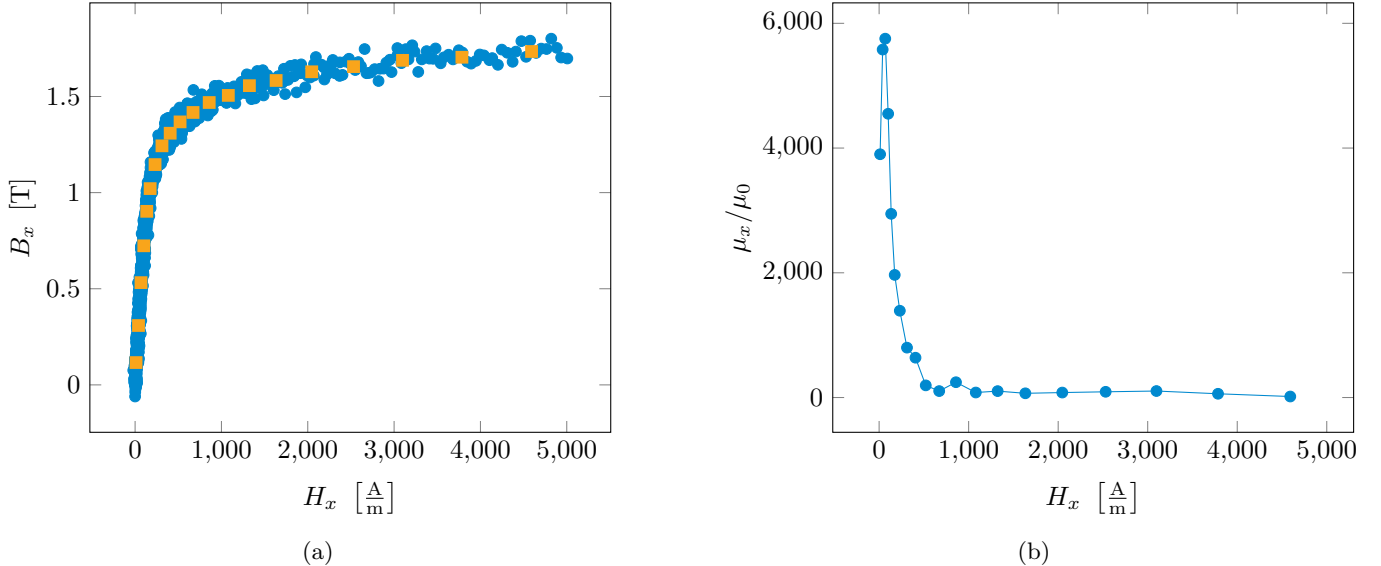
\begin{figure}[t!]
	\begin{subfigure}[t]{0.47\textwidth}
	\begin{tikzpicture}
	\centering
	\begin{axis}[ylabel shift = -3pt, xlabel={$H_x\,\,\left[\frac{\mathrm{A}}{\mathrm{m}}\right]$},ylabel={$B_x\,\,\left[\mathrm{T}\right]$}]
	\addplot[mark=*, only marks, TUDa-2b] table[x index=0, y index=1]{figures/BHnoisy.txt};
	\addplot[mark=square*, only marks, thick, TUDa-7b] table[x index=0, y index=1]{figures/BHcluster.txt};
	\end{axis}
	\end{tikzpicture}
	\caption{} 
	\label{fig:BH_noisy}
	\end{subfigure}
	\hfill
	\begin{subfigure}[t]{0.47\textwidth}
	\begin{tikzpicture}
	\centering
	\begin{axis}[ylabel shift = -3pt,xmax=5300, xlabel={$H_x\,\,\left[\frac{\mathrm{A}}{\mathrm{m}}\right]$},ylabel={$\mu_x/\mu_0$}]
	\addplot[mark=*, TUDa-2b] table[x index=0, y expr=\thisrowno{1}/(4*pi*1e-7) ]{figures/BHmus.txt};
	\end{axis}
	\end{tikzpicture}
	\caption{}
	\label{fig:noisy_mus}
	\end{subfigure}	
	\caption{(a) Noisy $BH$-data set in blue. K-means clusters in orange. (b) Estimated $\mu_x$, allocated at K-means cluster centers.}
	\label{fig:noisy}	
\end{figure}

Depending on the size of our measurement data set $\mathcal{D}$, we create $K \in \mathbb{N}$ clusters, such that the $(H,B)$-pair in each cluster share a ``similar'' differential reluctivity. 
To obtain the clusters, well-known K-means algorithms can be employed \cite{arthur2007, Bloemer2016, kaufmann1987,lloyd1982}.
The reluctivity of each cluster is then approximated with the linear coefficient of a standard linear regression model. Figure~\ref{fig:BH_noisy} shows exemplarily a measurement set with noisy data in blue and the clusters in orange. 
The corresponding estimation of the permeability $\mu$ is shown in Figure~\ref{fig:noisy_mus}.
To estimate the differential reluctivity or permeability of each cluster, a regression technique that is robust to outliers is needed. 
To that end, we employ Huber regression \cite{huber2009}. 
In contrast to conventional regression, points are classified as inliers and outliers and outlier points receive a linear weight that reduces their importance.
We note that, depending on the size and quality of the measurement data set, nonphysical solutions for the local permeability can occur, which will then have to be corrected \agcomment{according to \eqref{eq:mu_constraints}}.
\agcomment{Alternatively, one can employ manifolds learned on the measurement data \cite{kanno2020kernel, ibanez2018manifold}. 
These manifolds can then be employed to determine the local weighting factors in the data-driven iterations.}

Particularly for noisy data, we can take advantage of the fact that the weighting factors $\widetilde{\boldsymbol{\mu}}$ and $\widetilde{\boldsymbol{\nu}}$ do not necessarily have physical values in the data-driven context. 
Instead, the factors serve as indicators for the operating regions of each element.
Hence, in the noisy case, the data-driven algorithm utilizing local weighting factors is identical to the structure of Algorithm~\ref{alg:local_dd}, meaning that the manipulation of $\widetilde{\boldsymbol{\mu}}$ and $\widetilde{\boldsymbol{\nu}}$ is embedded in the maximum-entropy data-driven algorithm \cite{kirchdoerfer2017data}. Algorithm~\ref{alg:local_dd_noisy} shows the embedding of the local weighting factors assignment into the maximum entropy data-driven solver.

\begin{center}
	\begin{algorithm}[h!]
		\begin{algorithmic}
		\State \textbf{initialize} cluster centers $(\bar{\mathbf{H}}^\times, \bar{\mathbf{B}}^\times)$ and $\beta$ \cite{kirchdoerfer2017data} 
		\State \textbf{estimate} $\widetilde{\boldsymbol{\nu}}$ on the clustered measurement set $\mathcal{D}$ 
		\While{convergence not reached}
			\vspace{0.5em}
			\State Calculate weighted centers $(\bar{\mathbf{H}}^\times, \bar{\mathbf{B}}^\times)$ \cite{kirchdoerfer2017data} 
			\State Find $(\mathbf{H}^\times, \mathbf{B}^\times)$ in $\mathcal{D}$ adjacent to $(\bar{\mathbf{H}}^\times, \bar{\mathbf{B}}^\times)$ with distance function \eqref{eq:distance} 
			\State Solve variational problem \eqref{eq:discrete_problem_definition} for $\mathbf{a}$ and $\boldsymbol{\eta}$ 
			\vspace{0.25em}
			\State Update $\vec{B}_h =\mathrm{curl}\vec{a}_h$ 
			\State \phantom{Update }$\vec{H}_h = \vec{H}_h^\times +\widetilde{\boldsymbol{\nu}}\mathrm{curl} \vec{\eta}_h$ 
			\State Update $\beta$ according to \cite{kirchdoerfer2017data} 
			\vspace{0.5em}
			\If{\tikzmark{startb}stagnation detected}
				\State assign local reluctivity $\widehat{\boldsymbol{\nu}}$ to elements 
				\State assemble stiffness matrix $\mathbf{K}_{\hat{\nu}}$ 
				\State update distance function \eqref{eq:distance} with local reluctivity $\widehat{\boldsymbol{\nu}}$\tikzmark{endb}
			\EndIf
		\EndWhile
				\tikz[remember picture,overlay,pin distance=0cm]
		{\draw[line width=1pt,draw=black,fill=black,rectangle,rounded corners,fill opacity=0.1]
			([xshift=-15pt,yshift=10pt]$ (pic cs:startb)$ ) rectangle ([xshift=5pt,yshift=-16pt]$ (pic cs:endb)$ );
			\draw [decorate,decoration={brace,amplitude=10pt, mirror}]
			( $ (pic cs:endb) + (1ex,-3.7ex) $ ) --  ( $ (pic cs:endb) + (1ex,11.2ex) $ )  node  [align=left,xshift=2ex,black,right,pos=0.5] 
			{\emph{only for local weighting} \\ \emph{factor assignment}};
		}
		\end{algorithmic}
		\caption{Maximum entropy data-driven magnetostatic field solver. The gray part shows the extension according to the local weighting factors.}  
		\label{alg:local_dd_noisy}
	\end{algorithm}
\end{center}

\subsection{Numerical experiments}
\label{subsec:results_noisy}
The model under consideration is again the quadrupole magnet presented in Section~\ref{subsec:noisefree_numerical_experiments}. 
The maximum-entropy data-driven solver is applied both with global and local weighting factors. 
The noisy measurement data sets are generated by sampling the material curves similar to the noiseless case. 
The measurement data are then polluted with additive Gaussian noise according to formula \eqref{eq:noise}, \agcomment{thus, the equidistance property is now lost for both the $x-$ and $y$-components of the material property.} The chosen variances are $\sigma_{H}=10 \: \mathrm{A/m}$ and $\sigma_{B}=0.04 \: \mathrm{T}$. 
Figure~\ref{fig:BH_noisy} shows exemplarily the noisy measurement set in $x$-direction for $N=10^3$ data points.
Similar to the numerical investigations of Section~\ref{sec:noisefree}, we monitor the convergence of the data-driven solvers for data sets of increasing size using the errors \eqref{eq:err_em}, \eqref{eq:err_rms_H}, and \eqref{eq:err_rms_B}. 
Additionally, to estimate the variation of the data-driven solutions due to noise and assess the robustness of each data-driven solver, $100$ different noisy data sets are generated for each data set size.

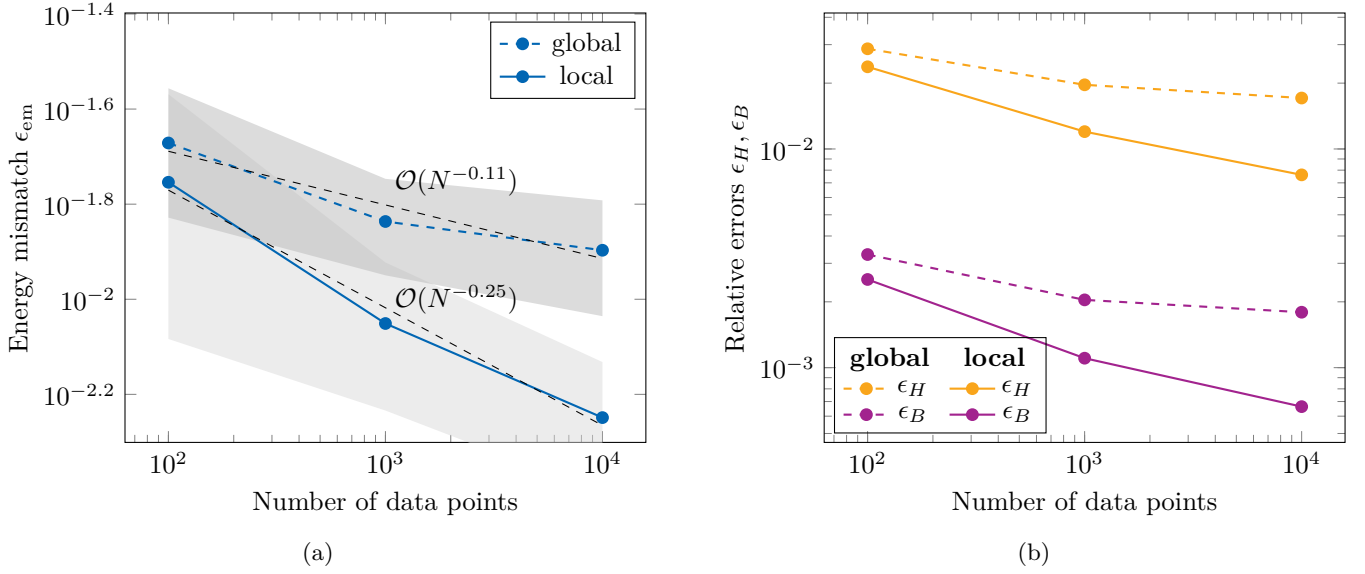
\begin{figure}[t]
	\begin{subfigure}[t]{0.47\textwidth}
		\begin{tikzpicture}
		\centering
		\begin{axis}[ylabel shift = -3pt,ymax=4e-02,ymin=5e-03, xmode=log,ymode=log,xlabel={Number of data points},ylabel={Energy mismatch $\epsilon_\text{em}$}]
		\addplot[name path=mean_dist_global,mark=*,  mark options={solid}, thick, dashed,TUDa-1b] table[x=N, y=dist_mean, col 	sep=comma]{figures/quadrupole/noisy/global/global_noisy_means.txt};
		\addlegendentry{global}
		
		\addplot[name path=mean_dist_local, mark=*, thick, TUDa-1b] table[x=N, y=dist_mean, col sep=comma]{figures/quadrupole/noisy/local/local_noisy_means.txt};
		\addlegendentry{local}
		
		\addplot[dashed, black] table[x=N, y=dist_least_square, col sep=comma]{figures/quadrupole/noisy/local/local_noisy_means.txt};
		\node[color=black,scale=1, anchor=west] at (axis cs:1e+3, 9.997226714587886272e-03) {$\mathcal{O}(N^{-0.25})$};
		
		\addplot[dashed, black] table[x=N, y=dist_least_square, col sep=comma]{figures/quadrupole/noisy/global/global_noisy_means.txt};
		\node[color=black,scale=1, anchor=west] at (axis cs:1e+3, 1.768693010036948466e-02) {$\mathcal{O}(N^{-0.11})$};
		
		\addplot[name path=var_dist_local_p,draw=none] table[x=N, y expr=\thisrowno{5} + 3* \thisrowno{6} , col sep=comma]{figures/quadrupole/noisy/local/local_noisy_means.txt};
		\addplot[name path=var_dist_local_m,draw=none] table[x=N, y expr=\thisrowno{5} - 3* \thisrowno{6} , col sep=comma]{figures/quadrupole/noisy/local/local_noisy_means.txt};	
		
		\addplot[gray!30,opacity=0.5]fill between[of=mean_dist_local and var_dist_local_p, soft clip={domain=-2.19:2.19}];%
		\addplot[gray!30,opacity=0.5]fill between[of=mean_dist_local and var_dist_local_m, soft clip={domain=-2.19:2.19}];
		
		\addplot[name path=var_dist_global_p,draw=none] table[x=N, y expr=\thisrowno{5} + 3* \thisrowno{6} , col sep=comma]{figures/quadrupole/noisy/global/global_noisy_means.txt};
		\addplot[name path=var_dist_global_m,draw=none] table[x=N, y expr=\thisrowno{5} - 3* \thisrowno{6} , col sep=comma]{figures/quadrupole/noisy/global/global_noisy_means.txt};
		
		\addplot[gray!55,opacity=0.5]fill between[of=mean_dist_global and var_dist_global_p, soft clip={domain=-2.19:2.19}];
		\addplot[gray!55,opacity=0.5]fill between[of=mean_dist_global and var_dist_global_m, soft clip={domain=-2.19:2.19}];	
		\end{axis}
		\end{tikzpicture}
		\caption{}
		\label{fig:quad_conv_distance_noisy}
	\end{subfigure}
	\hfill
	\begin{subfigure}[t]{0.47\textwidth}
		\begin{tikzpicture}
		\centering
		\begin{axis}[legend columns=2, legend style={/tikz/column 2/.style={column sep=5pt}},ylabel shift = -3pt, xmode=log, ymode=log, xlabel={Number of data points}, ylabel={Relative errors $\epsilon_{H}, \epsilon_{B}$}, legend style={fill=none, at={(axis cs:70,0.0005)}, anchor=south west}]
			\addlegendimage{empty legend}
			\addlegendentry{\hspace{-.6cm}\textbf{global}}
			\addlegendimage{empty legend}
			\addlegendentry{\hspace{-.6cm}\textbf{local}}
		\addplot[mark=*, mark options={solid},thick, dashed,TUDa-7b] table[x=N, y=rms_H_rel_mean, col sep=comma]{figures/quadrupole/noisy/global/global_noisy_means.txt};
		\addlegendentry{$\epsilon_{H}$} 
		\addplot[mark=*, thick, TUDa-7b] table[x=N, y=rms_H_rel_mean, col sep=comma]{figures/quadrupole/noisy/local/local_noisy_means.txt};
		\addlegendentry{$\epsilon_{H}$} 
		\addplot[mark=*, mark options={solid}, thick, dashed,TUDa-10b] table[x=N, y=rms_B_rel_mean, col sep=comma]{figures/quadrupole/noisy/global/global_noisy_means.txt};
		\addlegendentry{$\epsilon_{B}$} 
		\addplot[mark=*, thick, TUDa-10b] table[x=N, y=rms_B_rel_mean, col sep=comma]{figures/quadrupole/noisy/local/local_noisy_means.txt};
		\addlegendentry{$\epsilon_{B}$} 
		\end{axis}
		\end{tikzpicture}
		\caption{}
		\label{fig:quad_rms_noisy}
	\end{subfigure}	
	\caption{Convergence of data-driven solutions with respect to data set size for noisy measurement data. The solid lines refer to the data-driven solver utilizing local weighting factors, while the dashed lines refer to the standard data-driven solver based on a global weighting factor. (a) Convergence of the average energy mismatch for increasing data set size. The shaded areas show $\pm 3\sigma$ around the average errors. (b) Convergence of the average relative errors in the $\vec{H}$- and $\vec{B}$-field for increasing data set size.}
	\label{fig:quad_noisy_conv}	
\end{figure}

Using data sets of increasing size, the convergence of the data-driven solutions with respect to the average energy mismatch, computed over the $100$ different data sets, is shown in 
Figure~\ref{fig:quad_conv_distance_noisy}. 
The same figure also depicts the interval $\pm 3 \sigma$ around the average energy mismatches (in gray).
The convergence with respect to the average relative field errors $\epsilon_{H}$ and $\epsilon_{B}$, respectively defined in \eqref{eq:err_rms_H} and \eqref{eq:err_rms_B}, is presented in Figure~\ref{fig:quad_rms_noisy}. 
It is obvious from both figures that the maximum-entropy data-driven solver modified with local weighting factors results in accuracy and efficiency gains, improving the convergence rate of the method by a factor slightly greater than $2$. 
Compared to the noiseless case, i.e. looking at the results of Section~\ref{subsec:noisefree_numerical_experiments}, the convergence of both solvers is affected significantly by the addition of noise in the data. 
The impact of noise on the convergence of data-driven solvers has already been documented for the case of global weighting factors \cite{kirchdoerfer2017data}. 
This impact is also present when local weighting factors are utilized.
Note that Kirchdoerfer and Ortiz presented a convergence order of $\mathcal{O}(N^{-0.22})$ for the case of noisy data \cite{kirchdoerfer2017data}. 
However, as already discussed in Section~\ref{sec:noisefree}, the convergence rate is here additionally affected by the strong nonlinearity in the response of the soft magnetic material \agcomment{and its inherently unbalanced data set}. 
Hence, the slower convergence rate is to be expected.

\begin{figure}[h!]
	\begin{subfigure}[t]{0.48\textwidth}
	\begin{tikzpicture}
	\centering
	\begin{axis}[legend columns=2, legend style={/tikz/column 2/.style={column sep=5pt}},ylabel shift = -5pt, xmode=linear,ymode=log,xlabel={Number of iterations},ylabel={Energy mismatch $\epsilon_\text{em}$}]
		\addlegendimage{empty legend}
		\addlegendentry{\hspace{-.6cm}\textbf{global}}
		\addlegendimage{empty legend}
		\addlegendentry{\hspace{-.6cm}\textbf{local}}
	\addplot[thick, TUDa-1b, dashed] table[x=iter, y=mean_dist_1e2, col sep=comma]{figures/quadrupole/noisy/global/global_noisy_iterations_distance.txt};
	\addlegendentry{$N=10^2$} 
	\addplot[thick, TUDa-1b] table[x=iter, y=mean_dist_1e2, col sep=comma]{figures/quadrupole/noisy/local/local_noisy_iterations_distance.txt};
	\addlegendentry{$N=10^2$} 
	\addplot[thick, TUDa-4b, dashed] table[x=iter, y=mean_dist_1e3, col sep=comma]{figures/quadrupole/noisy/global/global_noisy_iterations_distance.txt};
	\addlegendentry{$N=10^3$} 
	\addplot[thick, TUDa-4b] table[x=iter, y=mean_dist_1e3, col sep=comma]{figures/quadrupole/noisy/local/local_noisy_iterations_distance.txt};
	\addlegendentry{$N=10^3$} 
	\addplot[thick, TUDa-10b, dashed] table[x=iter, y=mean_dist_1e4, col sep=comma]{figures/quadrupole/noisy/global/global_noisy_iterations_distance.txt};
	\addlegendentry{$N=10^4$} 
	\addplot[thick, TUDa-10b] table[x=iter, y=mean_dist_1e4, col sep=comma]{figures/quadrupole/noisy/local/local_noisy_iterations_distance.txt};
	\addlegendentry{$N=10^4$} 
	\addplot[very thick, dashed, TUDa-9b] coordinates {
		(14.5,1)
		(14.5,8e-3)
	};	
	\node[color=TUDa-9b,scale=1, anchor=east, xshift = -2] at (axis cs:14.5, 0.02) {global};
	\node[color=TUDa-9b,scale=1, anchor=west, xshift = 2]  at (axis cs:14.5, 0.02) {local};
	\end{axis}
	\end{tikzpicture}
	\caption{}
	\label{fig:quad_conv_distance_over_iter_noisy}
	\end{subfigure}
	\hfill
	\begin{subfigure}[t]{0.48\textwidth}
	\begin{tikzpicture}
	\centering
	\begin{axis}[legend columns=2, legend style={/tikz/column 2/.style={column sep=5pt}},ylabel shift = -5pt, xmin=15,xmode=linear,ymode=log,xlabel={Number of iterations},ylabel={Energy mismatch $\epsilon_\text{em}$}]
		\addlegendimage{empty legend}
		\addlegendentry{\hspace{-.6cm}\textbf{global}}
		\addlegendimage{empty legend}
		\addlegendentry{\hspace{-.6cm}\textbf{local}}
	\addplot[name path=mean_dist_global_1e3, thick, TUDa-1b, dashed] table[x=iter, y=mean_dist_1e2, col sep=comma]{figures/quadrupole/noisy/global/global_noisy_iterations_distance.txt};
	\addlegendentry{$N=10^2$} 
	\addplot[name path=mean_dist_local_1e3, thick, TUDa-1b] table[x=iter, y=mean_dist_1e2, col sep=comma]{figures/quadrupole/noisy/local/local_noisy_iterations_distance.txt};
	\addlegendentry{$N=10^2$} 
	\addplot[name path=mean_dist_global, thick, TUDa-10b, dashed] table[x=iter, y=mean_dist_1e4, col sep=comma]{figures/quadrupole/noisy/global/global_noisy_iterations_distance.txt};
	\addlegendentry{$N=10^4$} 
	\addplot[name path=mean_dist_local, thick, TUDa-10b] table[x=iter, y=mean_dist_1e4, col sep=comma]{figures/quadrupole/noisy/local/local_noisy_iterations_distance.txt};
	\addlegendentry{$N=10^4$} 
	
	\addplot[name path=var_dist_local_p,draw=none] table[x=iter, y expr=\thisrowno{3} + 3* \thisrowno{6} , col sep=comma]{figures/quadrupole/noisy/local/local_noisy_iterations_distance.txt};
	\addplot[name path=var_dist_local_m,draw=none] table[x=iter, y expr=\thisrowno{3} - 3* \thisrowno{6} , col sep=comma]{figures/quadrupole/noisy/local/local_noisy_iterations_distance.txt};	
	
	\addplot[TUDa-10b!25,opacity=0.5]fill between[of=mean_dist_local and var_dist_local_p, soft clip={domain=-2.19:2.19}];%
	\addplot[TUDa-10b!25,opacity=0.5]fill between[of=mean_dist_local and var_dist_local_m, soft clip={domain=-2.19:2.19}];

	\addplot[name path=var_dist_global_p,draw=none] table[x=iter, y expr=\thisrowno{3} + 3* \thisrowno{6} , col sep=comma]{figures/quadrupole/noisy/global/global_noisy_iterations_distance.txt};
	\addplot[name path=var_dist_global_m,draw=none] table[x=iter, y expr=\thisrowno{3} - 3* \thisrowno{6} , col sep=comma]{figures/quadrupole/noisy/global/global_noisy_iterations_distance.txt};
	
	\addplot[TUDa-10b!50,opacity=0.5]fill between[of=mean_dist_global and var_dist_global_p, soft clip={domain=-2.19:2.19}];
	\addplot[TUDa-10b!50,opacity=0.5]fill between[of=mean_dist_global and var_dist_global_m, soft clip={domain=-2.19:2.19}];

	\addplot[name path=var_dist_local_p_1e3,draw=none] table[x=iter, y expr=\thisrowno{1} + 3* \thisrowno{4} , col sep=comma]{figures/quadrupole/noisy/local/local_noisy_iterations_distance.txt};
	\addplot[name path=var_dist_local_m_1e3,draw=none] table[x=iter, y expr=\thisrowno{1} - 3* \thisrowno{4} , col sep=comma]{figures/quadrupole/noisy/local/local_noisy_iterations_distance.txt};	
	
	\addplot[TUDa-1b!25,opacity=0.5]fill between[of=mean_dist_local_1e3 and var_dist_local_p_1e3, soft clip={domain=-2.19:2.19}];%
	\addplot[TUDa-1b!25,opacity=0.5]fill between[of=mean_dist_local_1e3 and var_dist_local_m_1e3, soft clip={domain=-2.19:2.19}];

	\addplot[name path=var_dist_global_p_1e3,draw=none] table[x=iter, y expr=\thisrowno{1} + 3* \thisrowno{4} , col sep=comma]{figures/quadrupole/noisy/global/global_noisy_iterations_distance.txt};
	\addplot[name path=var_dist_global_m_1e3,draw=none] table[x=iter, y expr=\thisrowno{1} - 3* \thisrowno{4} , col sep=comma]{figures/quadrupole/noisy/global/global_noisy_iterations_distance.txt};
	
	\addplot[TUDa-1b!50,opacity=0.5]fill between[of=mean_dist_global_1e3 and var_dist_global_p_1e3, soft clip={domain=-2.19:2.19}];
	\addplot[TUDa-1b!50,opacity=0.5]fill between[of=mean_dist_global_1e3 and var_dist_global_m_1e3, soft clip={domain=-2.19:2.19}];	
	
	\end{axis}
	\end{tikzpicture}
	\caption{}
	\label{fig:quad_conv_distance_over_iter_noisy_variance}
	\end{subfigure}	
	\caption{Convergence of data-driven solutions with respect to the number of solver iterations for noisy measurement data. The solid lines refer to the data-driven solver utilizing local weighting factors, while the dashed lines refer to the standard data-driven solver based on a global weighting factor. (a) Convergence of the energy mismatch for increasing solver iterations over noisy data sets of increasing size. (b) Convergence of the energy mismatch for increasing solver iterations over noisy data sets of increasing size. The shaded areas show $\pm 3\sigma$ around the average errors.}
	\label{fig:quad_noisy_over_iter}			
\end{figure}
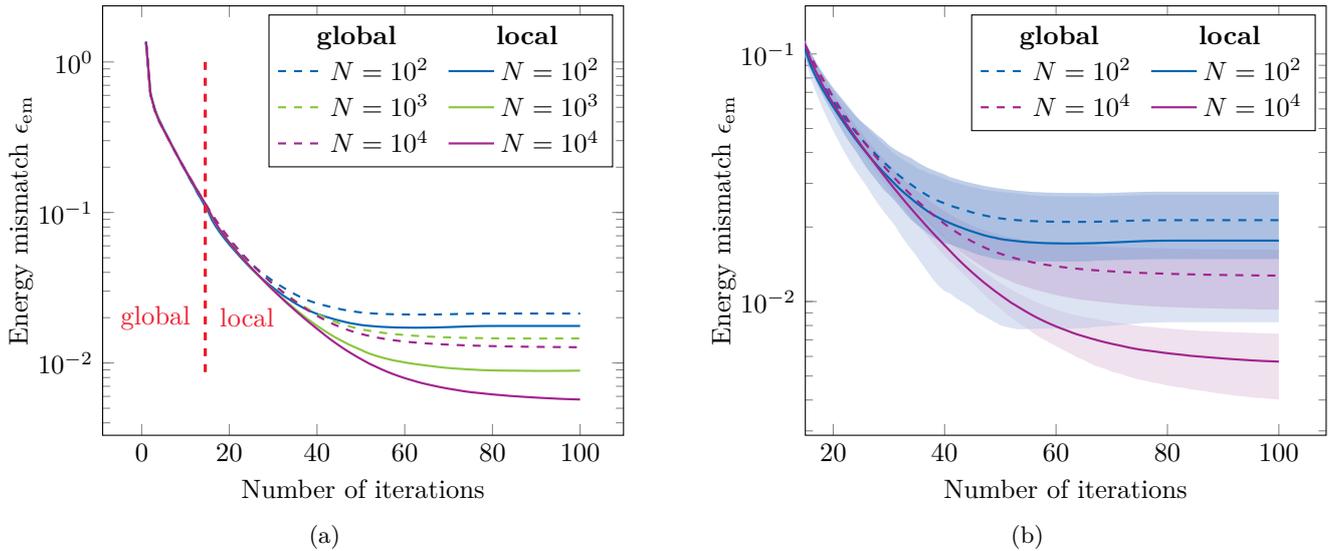

The convergence analysis with respect to the number of solver iterations is shown in Figures~\ref{fig:quad_conv_distance_over_iter_noisy} and \ref{fig:quad_conv_distance_over_iter_noisy_variance}.
Both figures show the average energy mismatch when global (dashed lines) and local (solid lines) weighting factors are utilized.
Figure~\ref{fig:quad_conv_distance_over_iter_noisy_variance} shows additionally the $\pm 3\sigma$ intervals around the average energy mismatch for measurements sets of size $N=10^2$ and $N=10^4$. 
Compared to the noiseless case, see Section~\ref{subsec:noisefree_numerical_experiments}, a substantial decrease in the number of iterations can be observed when the data-driven solver with a global weighting factor is employed.
On the contrary, when local weighting factors are used, the number of iterations is comparable to the noiseless case.
Thus, contrary to the noiseless case, assigning local weighting factors does not result in a significant improvement in terms of solver iterations. 
This difference can be attributed to the simulated annealing procedure, due to the fact that the solver iterations are now predominantly determined by controlling the constant $\beta$ \cite{kirchdoerfer2017data}.
 
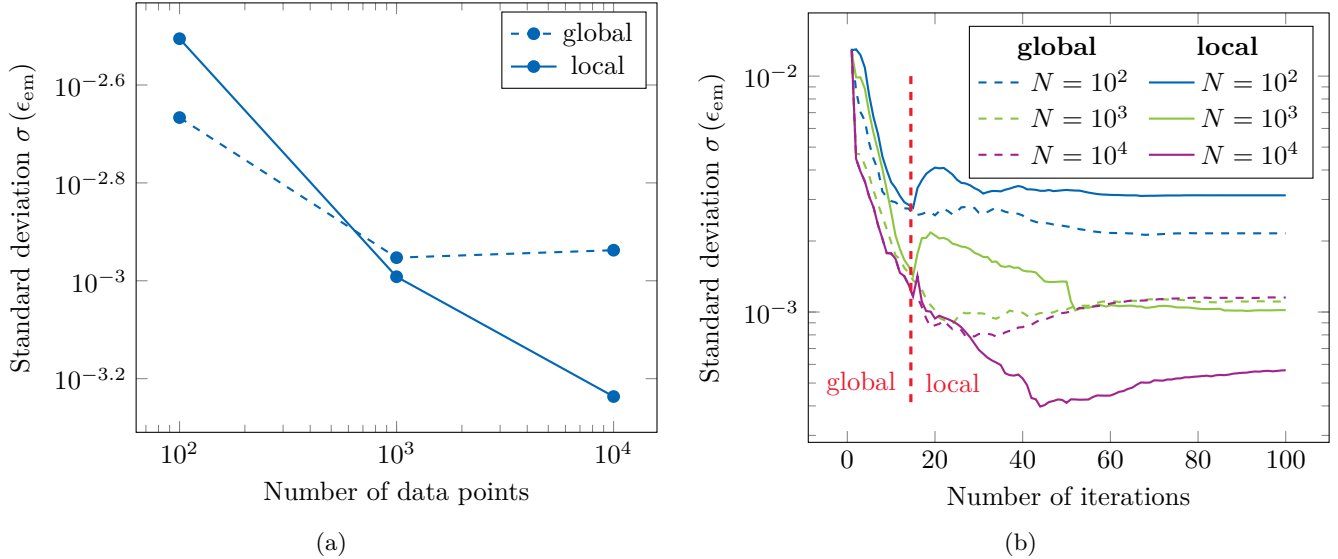
\begin{figure}[t]
	\begin{subfigure}[t]{0.49\textwidth}
	\begin{tikzpicture}
	\centering
	\begin{axis}[xmode=log,ymode=log,xlabel={Number of data points},ylabel={Standard deviation $\sigma \left( \epsilon_{\text{em}}\right)$}]
	\addplot[mark=*, mark options={solid}, thick, TUDa-1b, dashed] table[x=N, y=dist_var, col sep=comma]{figures/quadrupole/noisy/global/global_noisy_means.txt};
	\addlegendentry{global}
	\addplot[mark=*, thick, TUDa-1b] table[x=N, y=dist_var, col sep=comma]{figures/quadrupole/noisy/local/local_noisy_means.txt};
	\addlegendentry{local}
	\end{axis}
	\end{tikzpicture}
	\caption{}
	\label{fig:variance_number_of_dp}
	\end{subfigure}\hfill%
	\begin{subfigure}[t]{0.49\textwidth}
\begin{tikzpicture}
	\centering
	\begin{axis}[legend columns=2, legend style={/tikz/column 2/.style={column sep=5pt}},xmode=linear,ymode=log,xlabel={Number of iterations},ylabel={Standard deviation $\sigma \left( \epsilon_{\text{em}}\right)$}, legend pos=north east, legend style={fill=none}]
		\addlegendimage{empty legend}
		\addlegendentry{\hspace{-.6cm}\textbf{global}}
		\addlegendimage{empty legend}
		\addlegendentry{\hspace{-.6cm}\textbf{local}}
		\addplot[thick, TUDa-1b, dashed] table[x=iter, y=std_dist_1e2, col sep=comma]{figures/quadrupole/noisy/global/global_noisy_iterations_distance.txt};
		\addlegendentry{$N=10^2$} 
		\addplot[thick, TUDa-1b] table[x=iter, y=std_dist_1e2, col sep=comma]{figures/quadrupole/noisy/local/local_noisy_iterations_distance.txt};
		\addlegendentry{$N=10^2$} 
		\addplot[thick, TUDa-4b, dashed] table[x=iter, y=std_dist_1e3, col sep=comma]{figures/quadrupole/noisy/global/global_noisy_iterations_distance.txt};
		\addlegendentry{$N=10^3$} 
		\addplot[thick, TUDa-4b] table[x=iter, y=std_dist_1e3, col sep=comma]{figures/quadrupole/noisy/local/local_noisy_iterations_distance.txt};
		\addlegendentry{$N=10^3$} 
		\addplot[thick, TUDa-10b, dashed] table[x=iter, y=std_dist_1e4, col sep=comma]{figures/quadrupole/noisy/global/global_noisy_iterations_distance.txt};
		\addlegendentry{$N=10^4$} 
		\addplot[thick, TUDa-10b] table[x=iter, y=std_dist_1e4, col sep=comma]{figures/quadrupole/noisy/local/local_noisy_iterations_distance.txt};
		\addlegendentry{$N=10^4$} 
		\addplot[very thick, dashed, TUDa-9b] coordinates {
			(14.5,1e-2)
			(14.5,4e-4)
		};	
		\node[color=TUDa-9b,scale=1, anchor=east, xshift = -2] at (axis cs:14.5,5e-4) {global};
		\node[color=TUDa-9b,scale=1, anchor=west, xshift = 2]  at (axis cs:14.5,5e-4) {local};
	\end{axis}
\end{tikzpicture}
	\caption{}
	\label{fig:variance_iterations}
	\end{subfigure}	
	\caption{Standard deviation of the data-driven solutions for noisy measurement data. The solid lines refer to the data-driven solver utilizing local weighting factors, while the dashed lines refer to the standard data-driven solver based on a global weighting factor. (a) Standard deviation of the energy mismatch for noisy data sets of increasing size. (b) Standard deviation of the energy mismatch for increasing solver iterations over noisy data sets of increasing size.}
	\label{fig:variance_noisy}	
\end{figure}

The standard deviation $\sigma$ of the energy mismatch in dependence to the size of the data sets is shown in Figure~\ref{fig:variance_number_of_dp} and in dependence to the number of solver iterations in Figure~\ref{fig:variance_iterations}. 
Two main conclusions can be drawn from the presented results.
First, both figures show that the standard deviation decreases with larger data sets when local weighting factors are used. 
Thus, as should be expected, the robustness of the method is improved when larger data sets are available.
Contrarily, when a global weighting factor is used, no improvement in the variance is observed when increasing the data set size from $N=10^3$ to $N=10^4$. 
Second, for small data sets of size $N=10^2$, the standard deviation for the modified data-driven solver is slightly worse than the one of the standard solver.
Local weighting factors still offer an advantage for larger data sets and should be the method of choice in that case.
Nevertheless, in both cases, the differences are only marginal.

Overall, based on the variance improvement for larger data sets, as well as because of the convergence gains, local weighting factors are advantageous in the noisy case also.
However, the advantages are not as pronounced as in the noiseless case.

\section{\agcomment{Computational complexity}}
\label{sec:comp_complex}
\agcomment{The computational complexity of the data-driven Algorithm~\ref{alg:local_dd}, respectively Algorithm~\ref{alg:local_dd_noisy} for the noisy data case, is dictated by the cost for solving the two linear systems given in \eqref{eq:discrete_problem_definition}. 
Solving a linear system amounts to a complexity of $\mathcal{O}\left(N_\mathrm{e}^{3}\right)$ in the case of a standard direct solver. 
This is similar to the complexity of a standard Newton solver, the computational cost of which is also dominated by solving a linear system in each iteration. 
Next, we have to compare the number of iterations until the solvers reach a desired accuracy. 
In the case of the considered strongly nonlinear material, a rather small relaxation constant is necessary for the Newton solver. 
Both iterative solvers need approximately the same number of iterations for the same accuracy. 
Therefore, the overall cost of the data-driven solver is twice the cost of the Newton solver, since two linear systems must be solved in the former. 
Taking into account the computational complexity and the fact that it scales with the number of \glspl{dof}, we can conclude that if the Newton method is applicable to a given problem, then the data-driven solver shall be applicable as well, albeit at twice the computational cost. 
This conclusion is also valid in the case of computationally demanding three-dimensional magnetostatic field problems.}

\agcomment{We note that, if more than $10^5$ data points per dimension are employed, the computational cost of the discrete minimization problem is not anymore negligible, at least considering current implementations. Nevertheless, many strategies exist to improve the data-driven solver in that respect, e.g. constructing local neighbors around the field quantities at the quadrature point, such that only small parts of the measurement data set are employed in the minimization. Since data-driven solvers constitute a   young but very active research area, such improvements are to be expected in the near future.}

\section{Conclusion}
\label{sec:conclusion}
This work presented a data-driven, material-model-free, magnetostatic \gls{fem} solver, which addresses the challenging case of highly nonlinear material responses. 
The data-driven computing framework was extended with the introduction of local weighting factors, which indicate local operation points in the nonlinear domain. 
The novel approach was applied for the case of noiseless data by modifying an existing data-driven magnetostatics solver \cite{degersem2020magnetic}, based on the the original distance-minimization scheme \cite{kirchdoerfer2016data}. 
Likewise, for the case of noisy data, local weighting factors are embedded in the available maximum-entropy data-driven algorithm \cite{kirchdoerfer2017data}. 
The modified data-driven solvers were then compared to the standard data-driven solvers based on a global weighting factor.

The numerical results presented in Section~\ref{subsec:noisefree_numerical_experiments} for the noise-free case illustrate that the proposed modification to the data-driven algorithm leads to an impressive improvement in terms of convergence, accuracy, and computational efficiency. 
Therefore, in the case of noiseless data, local weighting factors should be preferred over the standard approach. 
\agcomment{Our work shows clearly that, for data-driven problems with a pronounced nonlinearity in the material behavior and a corresponding unbalanced data set, the use of local weighting factors is essential. This is also the case for balanced but generally sparse data sets coming from a nonlinear material response.}
The results also suggest that two data-driven \gls{em} solvers of increased intrusiveness which were introduced in a previous work \cite{degersem2020magnetic}, can be discarded altogether and replaced seamlessly by the approach presented in this paper.

In the case of noisy measurement data, the proposed modification of the maximum-entropy data-driven solver with local weighting factors improves the convergence rate of the solver by a factor of 2. 
The improvements in accuracy and efficiency become more pronounced for larger data sets, i.e. when more than $10^3$ data points are available.
Nevertheless, the addition of noise affects significantly the convergence behavior of both standard and modified data-driven solvers, therefore, the advantages due to local weighting factors are not as pronounced as in the noiseless case.
Similar conclusions are drawn when monitoring the variance of the data-driven solution, which is an indicator of the robustness of each data-driven method.
In particular, the data-driven solutions based on local weighting factors show a reduced variance for increasing data set size, whereas the global weighting factor-based solver seems to stagnate.
However, the variance differences are marginal, thus, the robustness of both methods can be seen as comparable.
Overall, the modification of the data-driven solver with local weighting factors is advantageous in the noisy case also, particularly if data sets with $N=10^3$ or more data points are available.

\subsection*{Acknowledgment}
D. Loukrezis and H. De Gersem would like to acknowledge the support of the Graduate School of Excellence for Computational Engineering at the Technische Universit\"at Darmstadt. D. Loukrezis further acknowledges the support of the BMBF via the research contract 05K19RDB. A. Galetzka's work is supported by the DFG through the Graduiertenkolleg 2128 ``Accelerator Science and Technology for Energy Recovery Linacs''.

\bibliographystyle{abbrv} 
\bibliography{references}

\textbf{}\end{document}